\documentclass{raa}            % referee version: for submission

\usepackage{graphicx,times}             %for PS/EPS graphics inclusion, new
\usepackage{natbib}
\usepackage{amssymb,amsmath}
\bibpunct{(}{)}{;}{a}{}{,}
\usepackage{lscape}

\usepackage[pagebackref=true]{hyperref}

\begin{document}

  \title{Structural Parameters of the Globular Cluster M\,15
}
%   \subtitle{I. Place Your Subtitle Here}

   \volnopage{Vol.0 (202x) No.0, 000--000}      %%preserved for Editor. DOn't remove!
   \setcounter{page}{1}          %%starting page, preserved for Editor. DOn't remove!

   \author{M. V. Petkova %% Put your Chinese name in "( )" if you like. 
      \inst{1,*}\footnotetext{$*$Corresponding Author}
   \and G. P. Petrov
      \inst{1}
      \and N. M. Kacharov
      \inst{2}
   \and P. L. Nedialkov
      \inst{1}%\footnotetext{ID: https:/orcid.org/0000-0002-0003-0004}
   }

   \institute{Faculty of Physics, Department of Astronomy, Sofia University `St. Kliment Ohridski',
5 J. Bourchier Blvd., Sofia 1164, Bulgaria; {\it mpetkova@phys.uni-sofia.bg}\\
%% Please give the E-mail address of the author, to whom future correspondence
%% requests will be sent.
        \and
             Leibniz Institute for Astrophysics (AIP), An der Sternwarte 16, 14482 Potsdam, Germany\\
        %\and
            % Full institute address for the third author\\
\vs\no
   {\small Received 2025 November 22; accepted 2026 March 12}}

\abstract{We present a detailed analysis of the structural parameters of the globular cluster M\,15 using g- and i-band photometric data from Pan-STARRS1 DR2. The central coordinates ($X_{\rm{C}}, Y_{\rm{C}}$), ellipticity ($\epsilon$), and position angle (PA) are derived via two independent methods: ellipse fitting of the two-dimensional stellar number isodensity distribution and Markov Chain Monte Carlo (MCMC) sampling. Our analysis of 38 stellar density intervals reveals a cluster center offset by only $4\,.\!\!^{\prime\prime}1\pm9\,.\!\!^{\prime\prime}6$ from the commonly accepted in the literature value, a good agreement on the order of our map resolution. We find a radial variation in the ellipticity, with a mean value of $\epsilon=0.09\pm0.02$ for the inner region ($R\leq4\,.\!\!^{\prime}5$) and $\epsilon=0.04\pm0.02$ for the outer region ($R>4\,.\!\!^{\prime}5$), where the errors correspond to $1\sigma$. The MCMC analysis of 75 datasets yields a mean $\epsilon=0.022\pm0.005$ for the entire cluster. The PA remains constant with increasing distance from the cluster center, $\rm{PA}=44\,.\!\!{\rm{^\circ}}4\pm16\,.\!\!{\rm{^\circ}}2$, and the MCMC method providing a consistent value of $\rm{PA}=46\,.\!\!{\rm{^\circ}}6\pm7\,.\!\!{\rm{^\circ}}1$. Our results are in agreement with some recent studies but challenge others, suggesting that a single $\epsilon$ value may be insufficient to fully characterize the overall oblateness of M\,15 due to incompleteness and crowding effects in its core. 
}

   \authorrunning{M. V. Petkova, G. P. Petrov, N. M. Kacharov \& P. L. Nedialkov}            %author_head in even pages
   \titlerunning{Structural Parameters of the Globular Cluster M\,15}  % title_head in odd pages

   \maketitle
%% The author head (on even pages) and the title head (on odd pages) will be
%% automatically extracted from \author{} and \title{}. Whenever the title is too long,
%% you will be asked to supply a shorter one by inserting either \authorrunning{} or
%% \titlerunning{} before \maketitle. Anyway, you can specify your own heads.
%%
%%
%% Note: In the following text body of your manuscript, please note several differences from
%%       other major journals:
%% (1) \subsection{Please Capitalize the First Letter of Each Notional Word in Subsection Title}
%% (2) Please Capitalize the First Letter of Each Notional Word in all tables' captions

%
%________________________________________________ sections below
%
\section{Introduction}           %% first-level sections will be auto-capitalized
\label{sect:intro}

Globular clusters (GCs) are gravitationally bound stellar systems, containing from tens of thousands to millions of stars. As some of the oldest objects in the Universe, they are of particular interest for understanding the early stages of galaxy formation and evolution. In the Milky Way, more than 200 GCs are known to exist (\citealt{Garro+etal+2024}), located predominantly in the halo, but also found in the bulge and the thick disk. A defining characteristic of GCs is their high stellar concentration, with densities reaching extreme levels toward their centers.
While often idealized as spherical, it has been known since the early 20th century that many GCs exhibit deviations from this shape, appearing as oblate spheroids (for a historical overview, see \citealt{Chen+Chen+2010}). This deviation is commonly quantified by the apparent ellipticity, $\epsilon=(1 - b/a)$, where a and b are the projected major and minor axes, respectively. The \cite{Harris+1996} catalog compiled comprehensive data for 157 GCs, providing ellipticity estimates for 100 of them. More recently, \cite{Reyes+Anderson+2024} listed the ellipticities for 163 Galactic GCs in their catalog.

M\,15 (NGC 7078) is one of the oldest known GCs in the Milky Way, with an estimated age of approximately 13 Gyr (\citealt{Baumgardt+etal+2023}) and a mass of 5.18$\,\times\,$$10^5\, \rm{{M}}_\odot$\footnote{\url{https://people.smp.uq.edu.au/HolgerBaumgardt/globular/}}. Historically, it is considered a prototypical core-collapsed cluster, characterized
by a surface density profile that rises steeply into the center -- a morphology
standing in marked contrast to the flat central cores of standard King-model
clusters (\citealt{Gerssen+etal+2002}). This central density cusp has long fueled the
debate over whether M\,15 harbors an intermediate-mass black hole (IMBH) or has
undergone a gravothermal core collapse. Most recently, \cite{Huang+etal+2025} provided
compelling kinematic evidence for an IMBH by identifying a hypervelocity star
ejected from the cluster’s center, effectively resolving this long-standing
dynamical puzzle.
Beyond its dense core, the external dynamics of M\,15 reveal a complex interaction
with the Galactic potential. Utilizing high-precision astrometry from Gaia DR3, \cite{Chen+etal+2025} recently detected faint tidal streams associated with the cluster,
allowing for a precise derivation of its mass loss rate. While this confirms that M\,15 is subject to active tidal stripping, it is important to note that at the smaller
radial scales probed in this work, these external tidal effects do not significantly
influence the cluster's intrinsic shape.
However, the geometric structure of the cluster -- specifically its ellipticity and
position angle -- remains a critical diagnostic for understanding its formation history
and dynamical state. In the context of the classification proposed by \cite{2024MNRAS.528.3198B}, M\,15 is a probable candidate for an accreted system, potentially
the stripped nucleus of a dwarf galaxy rather than an in-situ formed cluster. Such systems are expected to preserve internal rotation, which can result in a pronounced ellipticity in their inner regions (see, e.g., \citealt{Botev+etal+2025}). Therefore, precise measurements of geometric
parameters may provide crucial constraints on these evolutionary scenarios: a
detected inner ellipticity driven by rotation would support the stripped-nucleus
hypothesis, while a spherical morphology would be more consistent with a
non-rotating, pressure-supported system.

The main objective of this work is to derive the mean values of the structural parameters of M\,15 -- specifically its central coordinates ($X_{\rm{C}}, Y_{\rm{C}}$), ellipticity $\epsilon$, and position angle (PA) -- using photometric data from the Pan-STARRS1 (PS1) DR2\footnote{\url{https://outerspace.stsci.edu/display/PANSTARRS}} survey (\citealt{Flewelling+etal+2020}). These parameters are derived by applying ellipse fitting techniques and smoothing of two-dimensional (2D) density distribution at different angular scales.

This paper is organized as follows: Section 2 describes the data selection from the PS1 DR2 catalog and the filtering procedures applied. Section 3 details the two methods used to derive the structural parameters: ellipse fitting of the 2D number density distribution and discrete Markov Chain Monte Carlo (MCMC) sampling of the 2D number density distribution. The results are presented and compared with previous studies in Section 4. Finally, Section 5 provides a summary of the main conclusions of the work.

%% Authors can give a citation as 'Michel et al. 1992'.
%% You may also use \cite, \citep and \citet for citation, and use Table~1 or Figure~1
%% and so forth. Using \ref and \label for cross-references of Tables/Figures
%% is a good way in adjusting/adding/removing text, tables or figures.

\section{Data selection and filtering}
\label{sect:data}
We retrieved the gi-band photometry data from the PS1 DR2 catalog centered on the coordinates of M\,15: $\rm{RA}=21^{\rm {h}}$\,$ 29^{\rm {m}}$\,$58\,.\!\!\rm {^s}33$ and $\rm{DEC}=+12^{\circ}$\,$10'$\,$01\,.\!\!^{\prime\prime}2$ (J2000), as listed in \cite{Goldsbury+etal+2010}. 
The sources are required to have photometry in both the g- and i-bands, as this is necessary for subsequent analysis, such as filtering, constructing the color--magnitude diagram (CMD), etc. The availability of standard deviations for the photometry in both bands is also required, which will be used for the error assessment of our measurements.

To assess the impact of the field size on the obtained parameters, we downloaded the data for several angular scales.
The selection of the largest field of view was guided by the cluster's tidal radius, which is $27\,.\!\!^{\prime}3$ (\citealt{Harris+1996}). Consequently, we chose a $60^\prime$$\times$$60^\prime$ to ensure the entire cluster was encompassed. Subsequently, we also extracted smaller fields of $50^\prime$$\times$$50^\prime$, $40^\prime$$\times$$40^\prime$, $30^\prime$$\times$$30^\prime$, and $20^\prime$$\times$$20^\prime$. The rationale for using these smaller fields relates to the cluster's structure. M\,15 is noted in the Harris catalog as a core-collapsed cluster, characterized by a much smaller core and effective radii, relative to its tidal radius. Since the vast majority of member stars are concentrated in this central region, we assumed that the derived structural parameters should not differ significantly between the various field sizes.

At first, we applied a series of filtering steps to the data for each field size. We used quality flags (q) to distinguish between point-like and extended sources within the given g- or i-bands. To isolate stellar sources (point-like), a criterion of q $<$ 16777216 was applied. This filtering was necessary to verify the extent to which the presence of objects (e.g., blended stars, background galaxies) flagged as "Extended in this band" (q = 16777216) in PS1 DR2 catalog (\citealt{Flewelling+etal+2020}) affects the estimated structural parameters.

\begin{figure}[!htb]
\centering
\includegraphics[width=1\textwidth, angle=0]{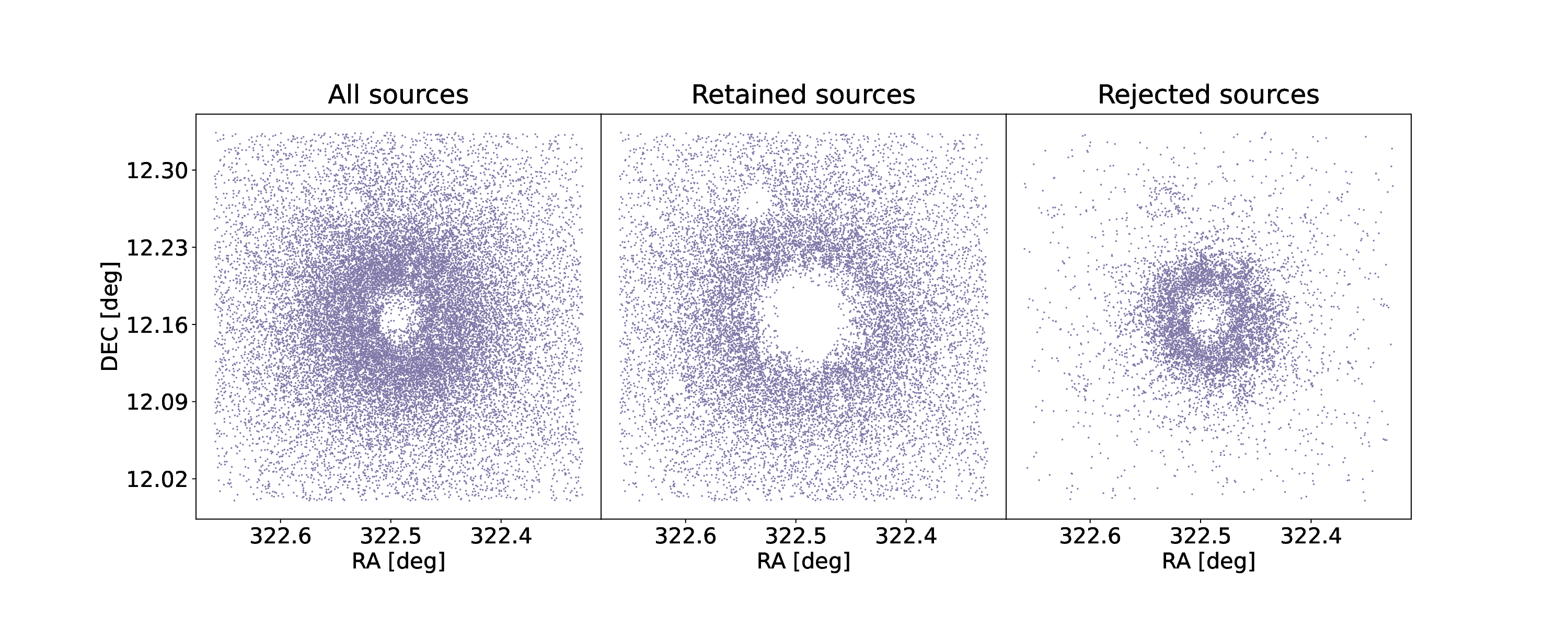}
\caption{Two-dimensional spatial distribution of the photometrically filtered stars of the GC M\,15 in a $20^\prime$$\times$$20^\prime$ field. The first panel from left to right shows the 2D distribution of the unfiltered sources, the second shows their distribution after filtering, and the third panel displays the 2D distribution of the rejected sources, which in this case satisfy the condition for the removal of extended objects in the g- or i-band (f4). As expected, these objects are predominantly located in the cluster's center and around the brightest stars.}
\label{filtered}
\end{figure}

\begin{figure}[!htb]
\centering
\includegraphics[width=1\textwidth, angle=0]{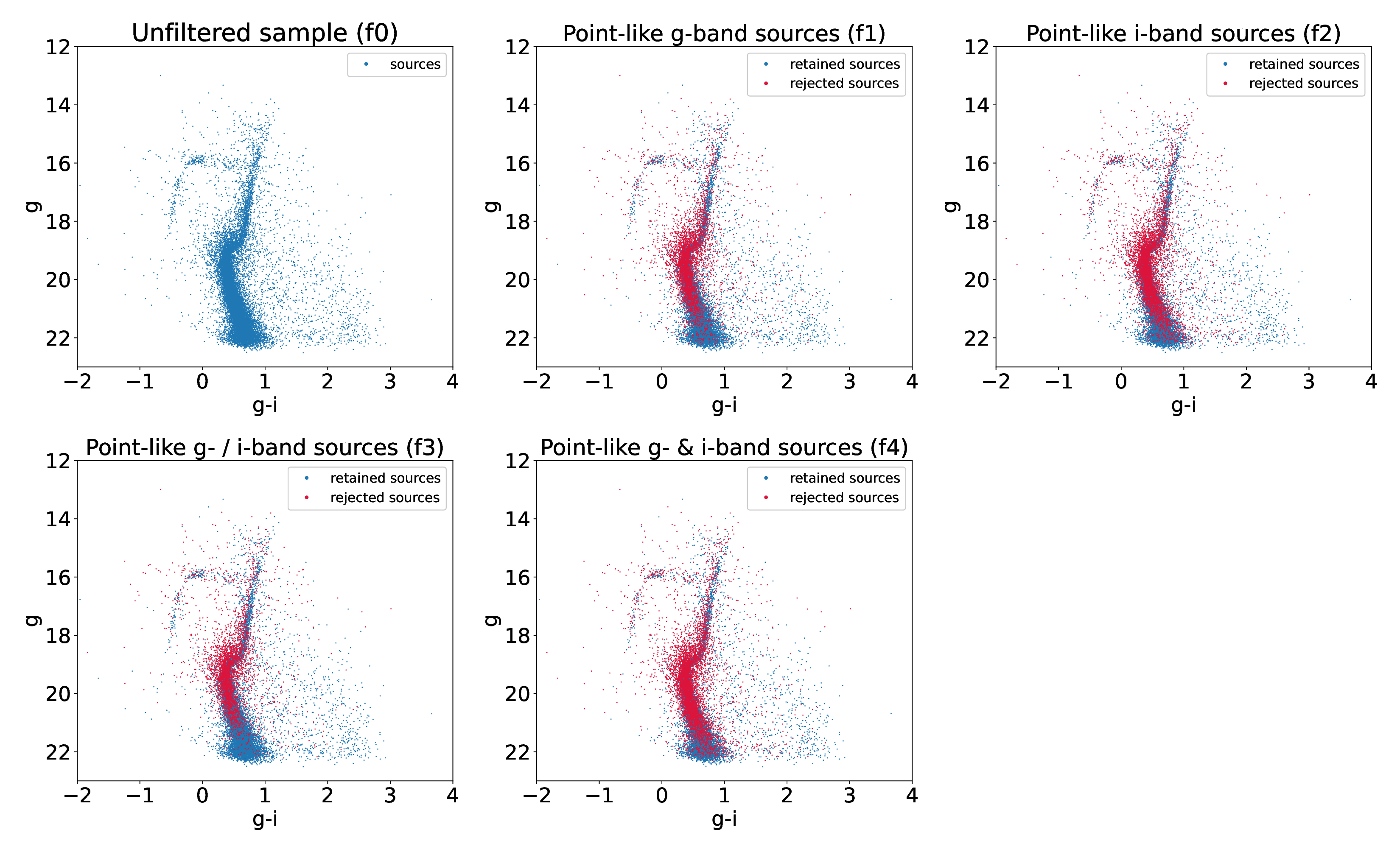}
\caption{CMDs of point-like sources (quality flag $<2^{24}=16777216$) in g- and i-bands within the 20$^\prime$$\times$20$^\prime$ field of M\,15. The top-left diagram shows the unfiltered sample with all sources plotted in blue. Each subsequent diagram presents one filtering type: retained objects are marked in blue and rejected ones with red. As expected, f4 excludes the largest fraction of sources.}
\label{filtered_cmd}
\end{figure}

We selected four types of filtration. The first is the removal of extended objects in the g-band (filtration f1). This left point-like objects in the g-band, while in the i-band, they can be either point-like or extended. We performed an analogous process for the i-band (filtration f2), which left point-like or extended sources in the g-band. The third filtration (f3) removed only the sources that are extended in both the g- and i-bands. In this case, sources that are point-like and those that are extended in only one of the two bands remained. The final case (f4) is the most precise filtration, that left only point-like sources in both g- and i-bands. An example of the filtration for a cluster in a $20^\prime$$\times$$20^\prime$ field is shown in Figure~\ref{filtered}, and the CMDs for each filtration type are illustrated in Figure~\ref{filtered_cmd}.

Following the filtering process, we transformed the equatorial coordinates (RA, DEC) of the sources to tangential coordinates (X, Y), using the M\,15 center coordinates given by \cite{Goldsbury+etal+2010}.

Finally, based on the methodology of \cite{{Lee+Carney+1999}}, we applied a contamination subtraction for the PS1 DR2 photometric data. For each field size, we selected two corresponding background fields. These are located $5^{\circ}$ from the cluster's center, but at the same Galactic latitude to ensure uniform background density. These areas have been examined and are devoid of any large, bright objects. They are intended to be used for the removal of background contamination from the GC, provided that the object density within them corresponds to the background object density surrounding the GC. Comparing the CMDs of M\,15 and the background field, upon matching with a field star within $3 \sigma$ of g and g--i, we excluded background objects from the M\,15 CMD.

Once each sample has been filtered and the background objects removed, the cluster's structural parameters can be derived.

\section{Methods}
\label{sect:methods}

\subsection{Structural parameters derived from ellipse fitting of the 2D number density distribution}
\label{iso}

Here we analyze the unfiltered 40$^\prime$$\times$40$^\prime$ field of view centered on M\,15, where the background contamination is accounted for, but the extended objects are not removed. We binned the data to generate the stellar number density, defined as the count of stars per pixel (*/px). We prefer stellar number density over the cluster's surface brightness, as it is less sensitive to individual bright stars or projected background contamination. Initially, the stellar density map was constructed using square pixels of $24^{\prime\prime}$$\times$$24^{\prime\prime}$ and $12^{\prime\prime}$$\times$$12^{\prime\prime}$, yielding arrays of 100$\times$100 and 200$\times$200 pixels, respectively. These pixel sizes were chosen as an optimal compromise. They are large enough to ensure sufficient star statistics per bin in the outskirts, yet small enough to resolve the structural changes in the inner regions without washing out the ellipticity signal.

After pixelizing the image, density thresholds can be applied directly to filter for a specific range of values, which is then used to derive the ellipse parameters. Alternatively, the map can be smoothed beforehand to reduce noise. Both approaches were tested for comparison. In the first case (unsmoothed), the density values are discrete, as they correspond to the integer number of stars in each pixel. In the second case, however, smoothing the cluster map resulted in continuous density values, which permitted a more refined selection of the density intervals.

\label{AppB}
We performed the smoothing using a 2D boxcar averaging method. This was implemented with a 3$\times$3 or 5$\times$5 kernel where each element has a weight of 1, while all elements outside this matrix have a weight of 0. A thorough validation of this smoothing procedure using mock density maps is presented in Appendix~\ref{AppB}, confirming that it introduces no systematic errors to the derived structural parameters.

Following the smoothing procedure, we selected specific density intervals for analysis. These intervals were chosen to avoid regions very close to the cluster's center, where data incompleteness can lead to spurious results. Similarly, we excluded the outermost parts of the cluster, as the data there is noisier and the densities are lower.

\begin{figure}
\centering
\includegraphics[width=0.7\textwidth, angle=0]{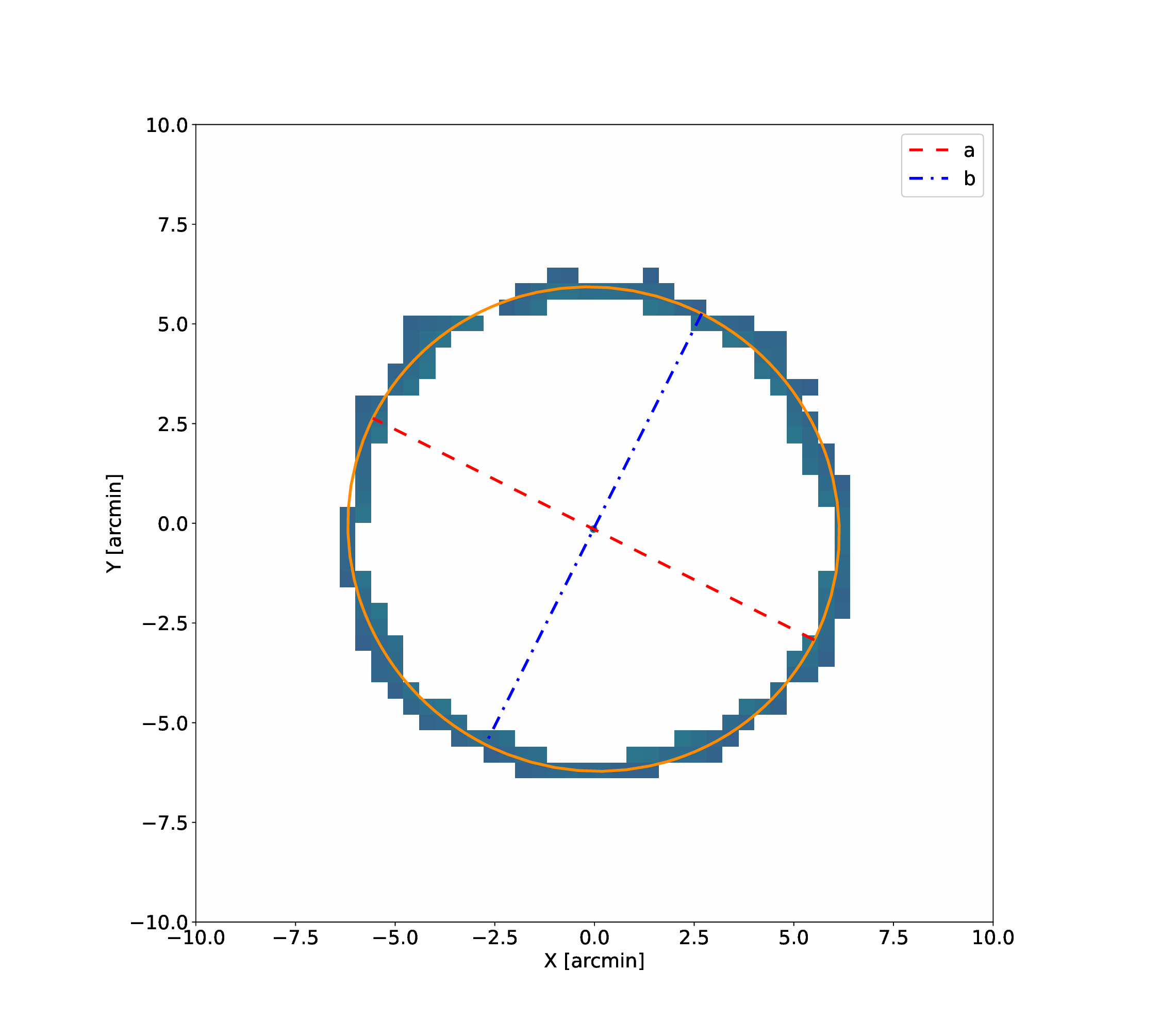}
\caption{Ellipse fitting over the 7--9 */px density interval for the 100$\times$100 array smoothed with a 5$\times$5 boxcar. The major and minor axes are indicated by a red dashed line and a blue dash-dotted line, respectively. The fitted ellipse is drawn in orange and the isodensities are marked with blue pixels.}
\label{ell_7-9}
\end{figure}

For each selected interval, we fitted an ellipse using the EllipseModel from the scikit-image Python package (\citealt{van der Walt+etal+2014}), which implements a least-squares algorithm. This fit yields the cluster's structural parameters: the semi-major $a$ and semi-minor $b$ axes, the center coordinates ($X_{\rm{C}}, Y_{\rm{C}}$), the position angle PA, and the ellipticity $\epsilon$. An example of an ellipse fit for the 7--9 */px density interval, applied to a 100$\times$100 array smoothed with a 5$\times$5 boxcar kernel, is shown in Figure~\ref{ell_7-9}.

We performed error estimation using the Bootstrap method (\citealt{Efron+1979}). This technique involves repeatedly generating new datasets by resampling the coordinates of the pixels within a given density interval. Crucially, this is done with replacement, meaning a single pixel's coordinates can be selected multiple times in a new sample. This process is repeated 10\,000 times, creating 10\,000 bootstrapped datasets. An ellipse is then fitted to each of these, yielding a distribution of values for each parameter. We adopted the standard deviation of this distribution as the 1$\sigma$ error for that parameter.

\begin{figure}
\centering
\includegraphics[width=1\textwidth, angle=0]{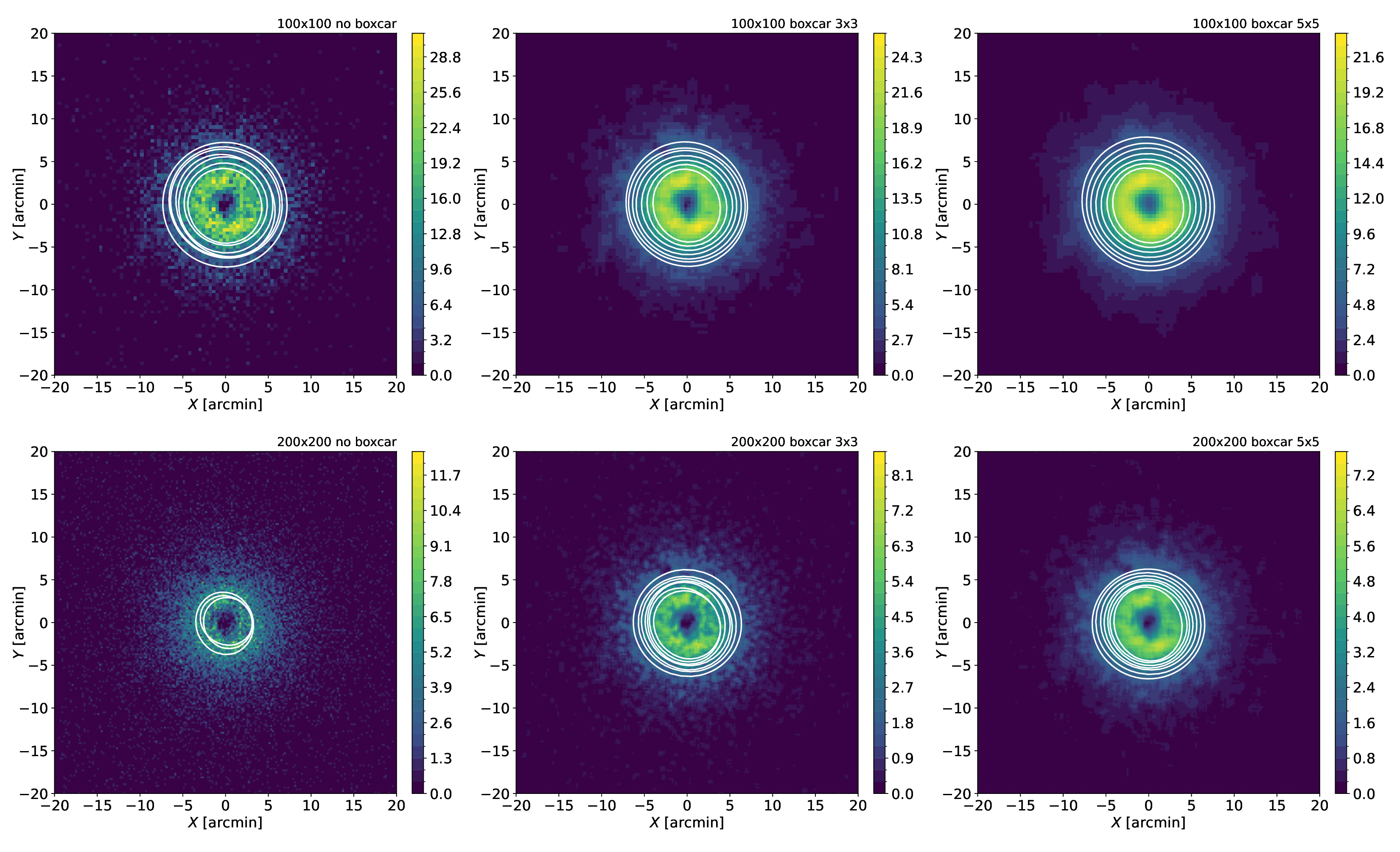}
\caption{Ellipse fitting results for all considered 38 stellar density intervals within the 40$^\prime$$\times$40$^\prime$ field. From left to right the first row presents the 100$\times$100 array under the three conditions -- unsmoothed, 3$\times$3 boxcar and 5$\times$5 boxcar. The second row shows the same arrangement for the 200$\times$200 array. The color scale on the right of the panels represents the density */px.}
\label{all_ell}
\end{figure}

The results from the ellipse fitting for all considered 38 density intervals in the $40^\prime$$\times$$40^\prime$ field are presented in Table~\ref{Tab1} and are shown graphically in Figure~\ref{all_ell}. Due to the contribution of extended objects (see Figure~\ref{filtered}) the highest stellar density is observed between 1$^\prime$ and 3$^\prime$ from the cluster center. Although a torus-like structure appears when 24$^{\prime\prime}$ binning is combined with 5$\times$5 boxcar smoothing, the densities within 3$^\prime$ remain patchier in all other cases. Consequently, we adopt an inner fitting radius of $\sim3$$^\prime$ for all samples in order to avoid the influence of extended sources.

\subsection{Structural parameters obtained via discrete MCMC sampling}
\label{MCMC}

We also derive the structural parameters of the GC M\,15 -- the central coordinates ($X_{\rm{C}}, Y_{\rm{C}}$), the PA and the ellipticity $\epsilon$ by fitting a 2D Plummer model, using a modified version of the discrete maximum likelihood method \footnote{\url{https://github.com/kacharov/morphology_2d}}. For more details, see \cite{Kacharov+etal+2014}. This method utilizes the discrete source distribution within the cluster and determines the most probable parameters through MCMC optimization.

The input data consist of the filtered samples for the various field sizes after the subtraction of one of the two background estimates, totaling 50 datasets (5 filters × 5 field sizes × 2 background fields). We analyzed additional 25 datasets (5 filters × 5 field sizes) without background subtraction. In total, 75 independent tests were conducted. The results are presented in Table~\ref{Tab2}, where column 2 presents the four types of filtration (f1--f4) and f0 designates unfiltered samples. The various types of background subtraction are listed in column 3: b1 and b2 indicate the two different background fields that were used for subtraction, b0 marks no background subtraction.

\begin{figure}
\centering
\includegraphics[width=0.7\textwidth, angle=0]{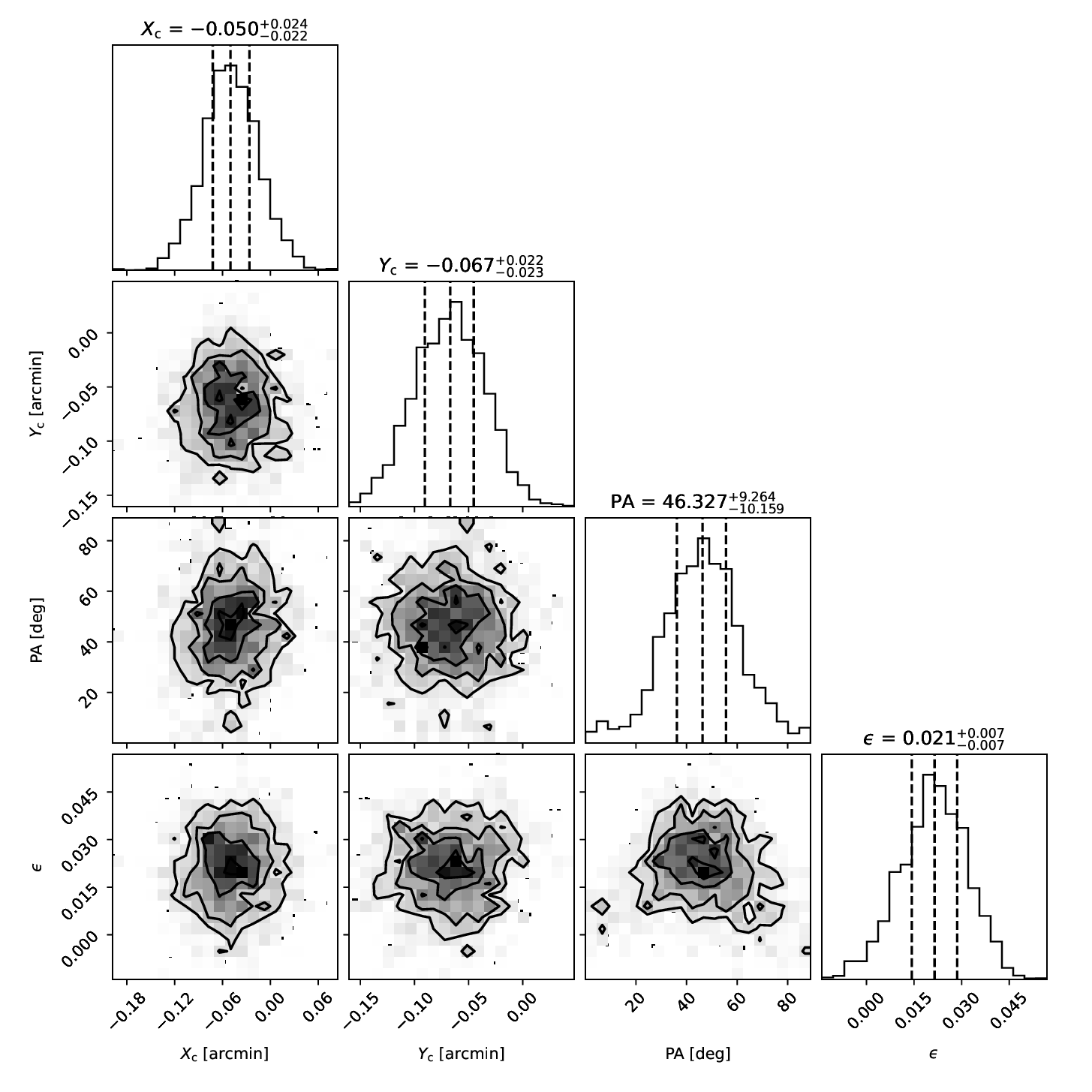}
\caption{Example of M\,15 structural parameters derived from MCMC for 50$^\prime$$\times$$50^\prime$ field using filtration f3 and background subtraction b2. The histograms show the distribution for each parameter, with vertical dashed lines marking the 0.25, 0.5, and 0.75 quantiles. The contour plots show the pairwise parameter correlations. The median values and corresponding errors are listed above each histogram.}
\label{corner_example}
\end{figure}

Our MCMC analysis used a characteristic radius of $r_{\rm{s}}=4\,.\!\!{^\prime}5$, chosen because interior to this point, both the sample completeness and photometric accuracy are significantly decreased by stellar crowding.
In this implementation, the python module {\sc emcee} 
MCMC algorithm (\citealt{Foreman-Mackey+etal+2013}) was initialized with 40 walkers for each parameter, then each of the 40 chains was run for 600 steps. The posterior distribution for each parameter was constructed from the values of the last 150 steps of each of the 40 walkers (6\,000 values total). The best-fit value for each parameter was taken as the median (the 0.5 quantile) of its distribution. The uncertainties were estimated from the 0.25 and 0.75 quantiles. Figure~\ref{corner_example} shows an example of a corner plot illustrating the correlations between the derived parameters, with contours representing the probability distributions.

\section{Results and discussion}
\label{sect:results}
In this section, we present our results for the structural parameters of the GC M\,15 and compare them with those proposed by other authors.

\begin{figure}
\centering
\includegraphics[width=0.7\textwidth, angle=0]{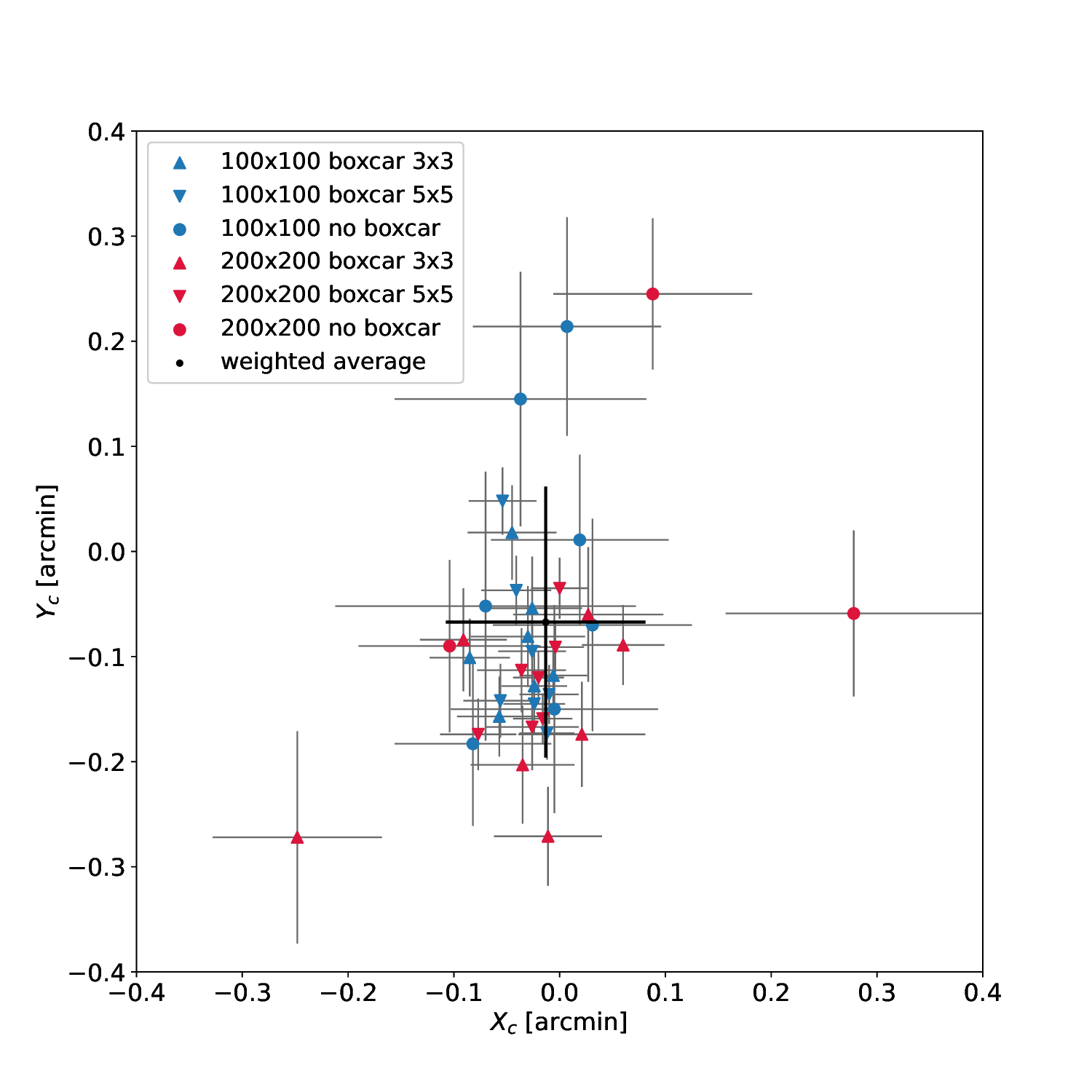}
\caption{M\,15 Center coordinates derived from 38 ellipses of different stellar densities with their weighted mean and standard deviation: $\Delta X_{\rm{C}}=-0\,.\!\!{^\prime}013\pm 0\,.\!\!{^\prime}094$ and $\Delta Y_{\rm{C}}=-0\,.\!\!{^\prime}067\pm0\,.\!\!{^\prime}129$, respectively. The mean is displaced by $\sim 0\,.\!\!{^\prime}068$ or $4\,.\!\!^{\prime\prime}1\pm 9\,.\!\!^{\prime\prime}6$ relative to the adopted center coordinates of M\,15 from \cite{Goldsbury+etal+2010}.}
\label{x_y_all}
\end{figure}

\begin{figure}
\centering
\includegraphics[width=1\textwidth, angle=0]{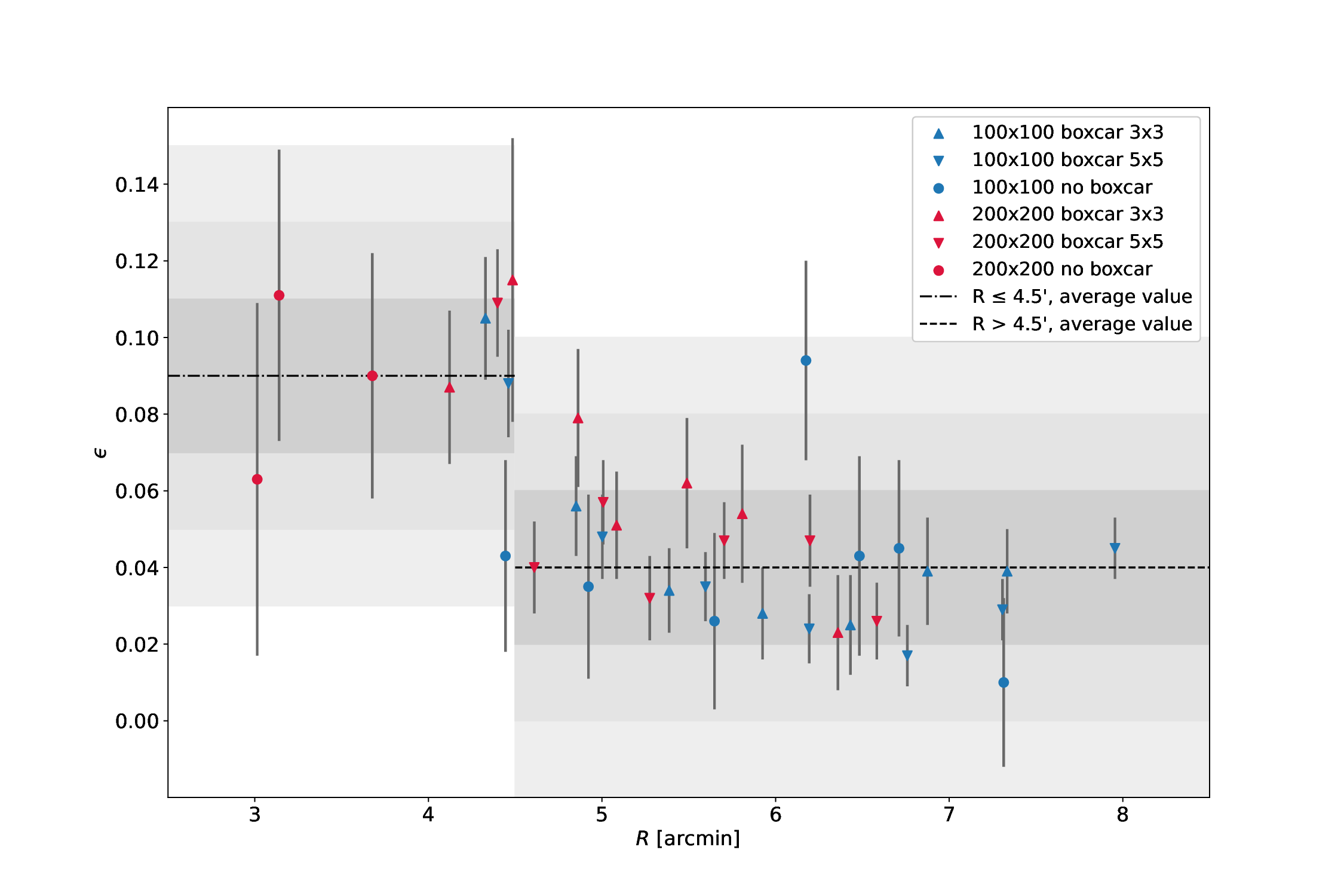}
\caption{Ellipticities $\epsilon$ of the 38 fitted ellipses at various radii. The mean $\epsilon$ values in the radial intervals $3\,.\!\!{^\prime}0<R \leq 4\,.\!\!{^\prime}5$ and $4\,.\!\!{^\prime}5<R<8\,.\!\!{^\prime}0$ are $ 0.09\pm0.02$ and $ 0.04\pm0.02$, respectively. The dark, medium, and light grey shaded regions correspond to the $\pm 1\sigma$, $\pm 2\sigma$, and $\pm 3\sigma$ intervals. }
\label{ell_a_all}
\end{figure}

\begin{figure}
\centering
\includegraphics[width=1\textwidth, angle=0]{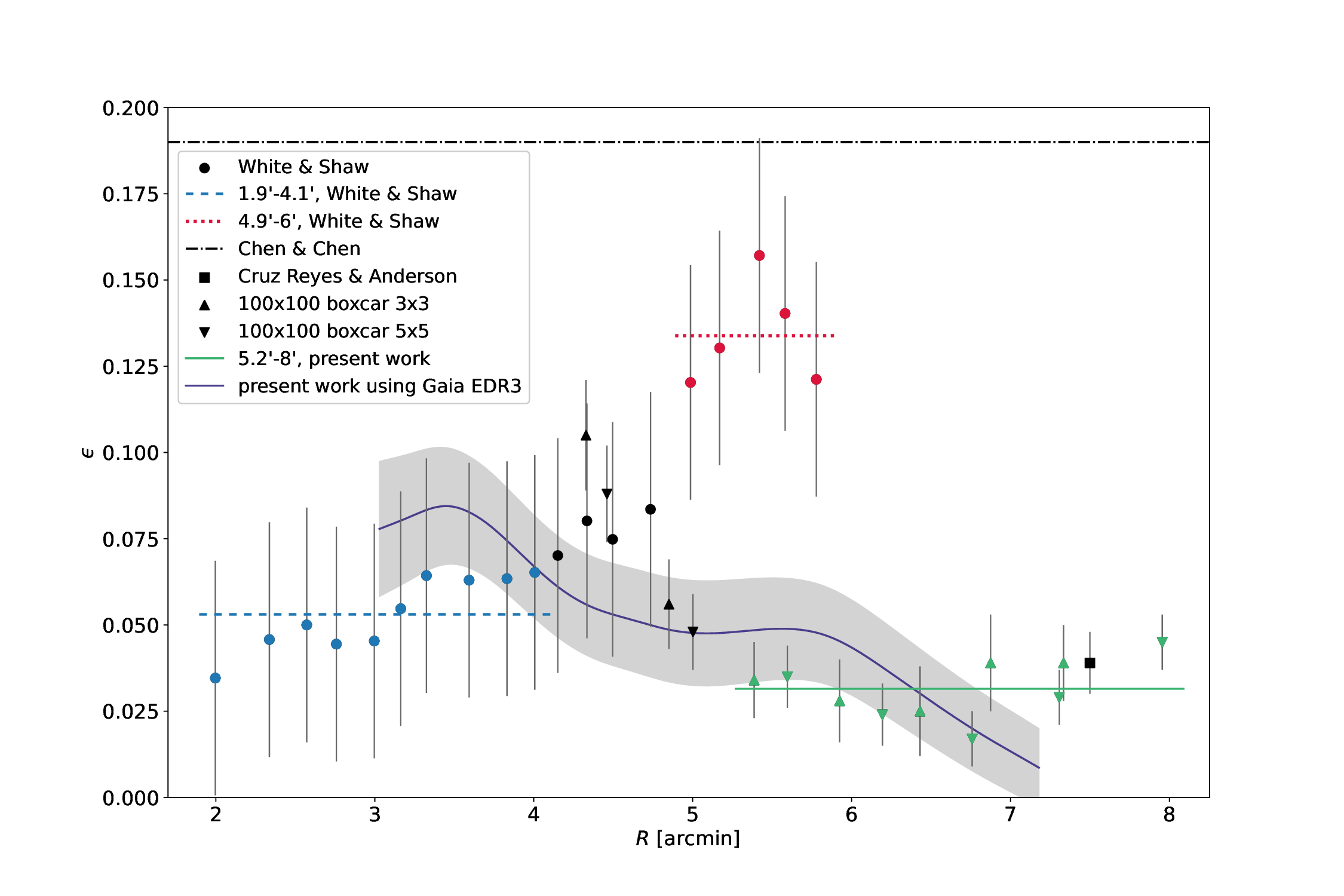}
\caption{Comparison of the ellipticity derived in this work with those reported in previous studies. The measurements of \cite{White+Shawl+1987} are denoted with dots; the blue dashed line marks their mean value of $0.05$ $\pm$ $0.01$ for radii between $1\,.\!\!{^\prime}9$ and $4\,.\!\!{^\prime}1$, and the red dotted line indicates $0.13$ $\pm$ $0.02$ for radii between $4\,.\!\!{^\prime}9$ and $6^\prime$. The data points used to compute these averages are shown in the corresponding colors. The mean ellipticity reported by \cite{Chen+Chen+2010} is shown as a black dash-dotted line and the black square represents the value measured by \cite{Reyes+Anderson+2024}. The triangles indicate our results based on PS1 DR2 data and the green solid line corresponds to the mean value of $0.03\pm0.01$ for the subset of our data highlighted in green. Alternatively, we perform 73 measurements by ellipse fitting of the smoothed number densities based on Gaia EDR3 data  (\citealt{Vasiliev+Baumgardt+2021}). The purple solid line and the grey shaded region show the variation of ellipticity within $\pm 1\sigma$ interval.}
\label{compare}
\end{figure}

\begin{figure}
\centering
\includegraphics[width=1\textwidth, angle=0]{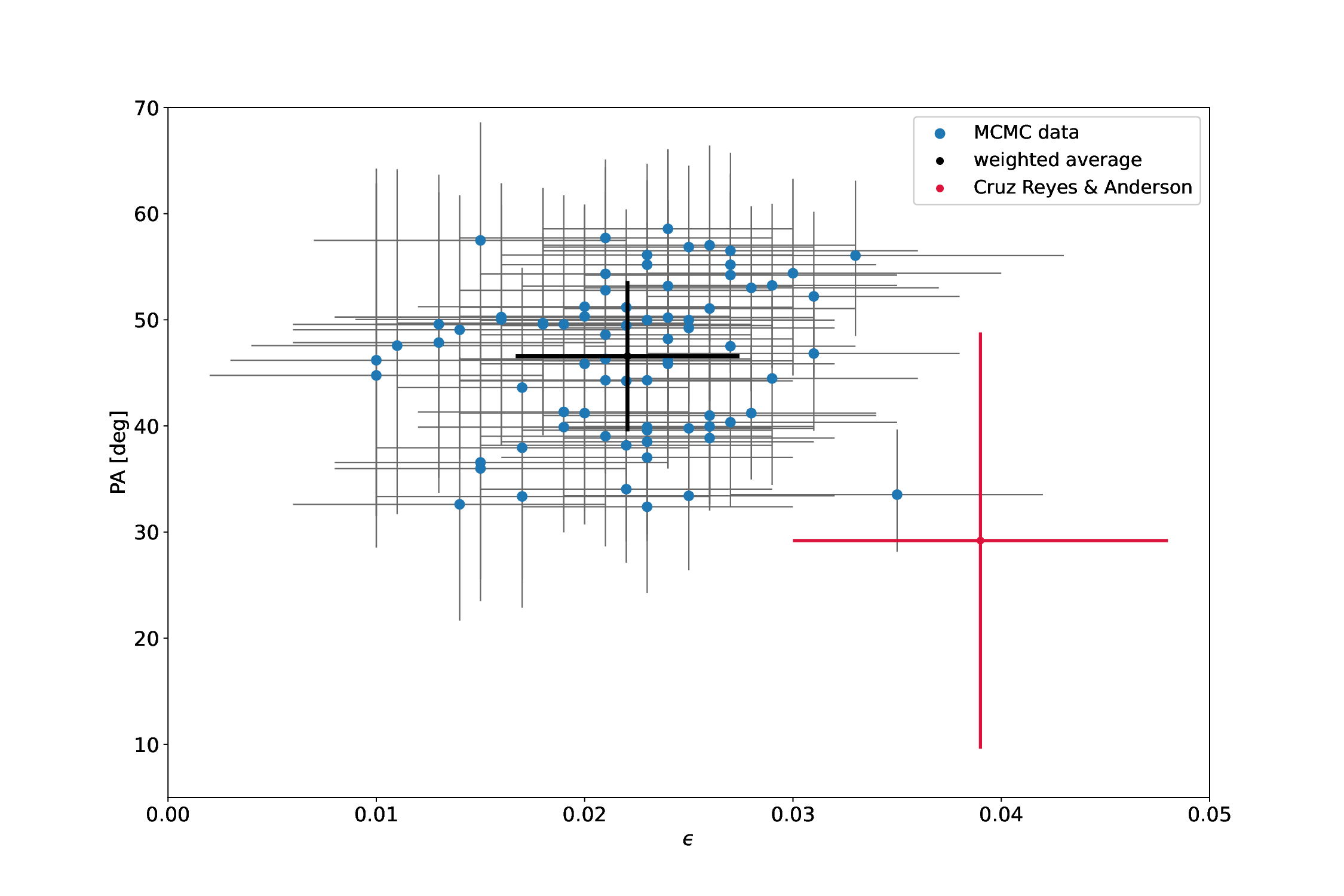}
\caption{Results of the 75 MCMC fits (see Table~\ref{Tab2}) to the structural parameters of M\,15 -- ellipticity (abscissa) and position angle (ordinate), and their weighted mean values and standard deviations: $\epsilon=0.022\pm0.005$ and $PA=46\,.\!\!{\rm{^\circ}}6\pm7\,.\!\!{\rm{^\circ}}1$, respectively.}
\label{all_ell_mcmc}
\end{figure}

\begin{figure}
\centering
\includegraphics[width=1\textwidth, angle=0]{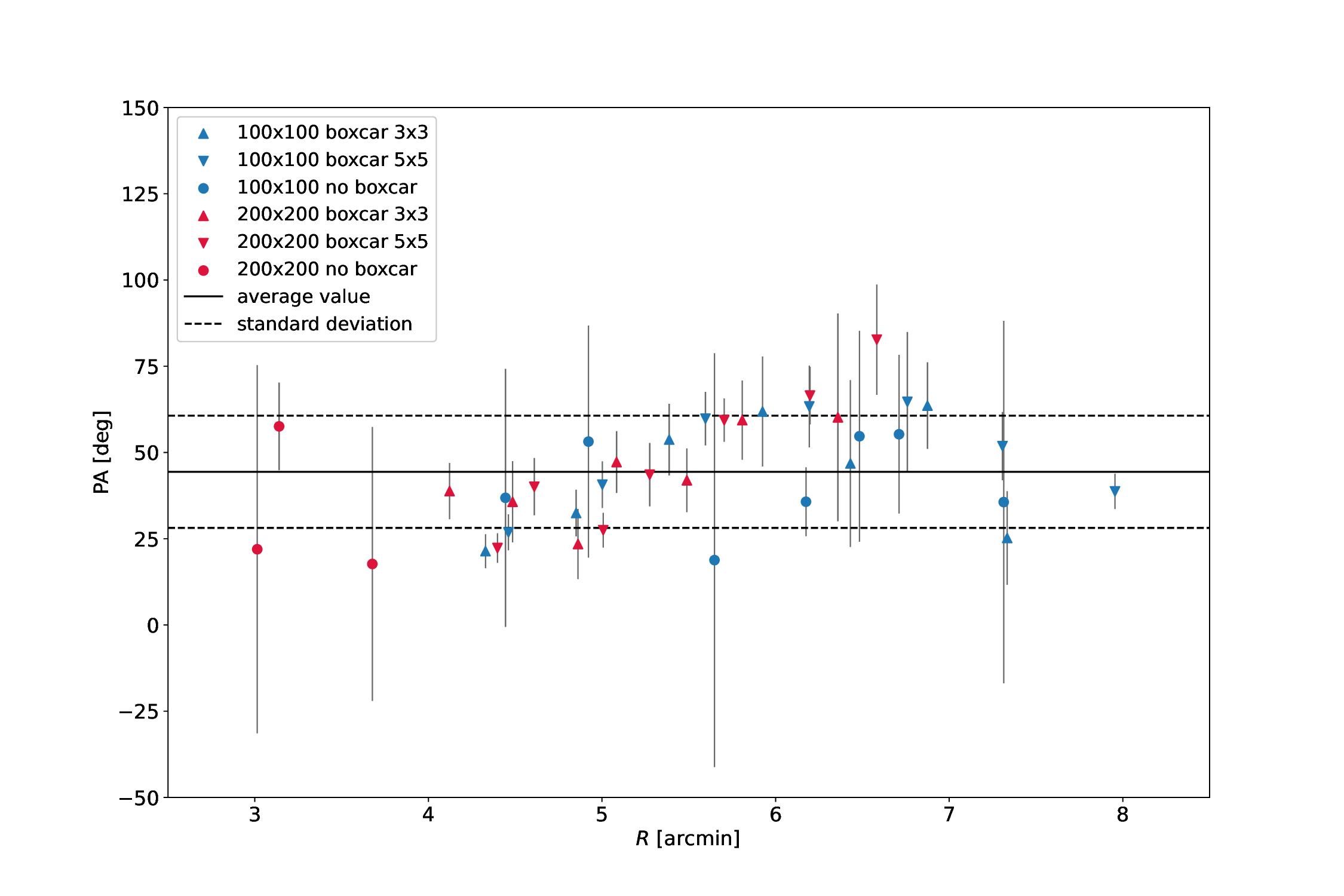}
\caption{Position angles of the 38 fitted ellipses with a mean value marked by a solid black line and a standard deviation indicated by dashed lines: PA = $44\,.\!\!{\rm{^\circ}}4\pm16\,.\!\!{\rm{^\circ}}2$. }
\label{pa_a_all}
\end{figure}

Figure~\ref{x_y_all} illustrates the agreement between the cluster center coordinates det 
ermined in our work and these from \cite{Goldsbury+etal+2010}. The figure shows the offsets $\Delta X_{\rm{C}}$ and $\Delta Y_{\rm{C}}$ (in arcmin) for the centers of 38 ellipses fitted to the smoothed stellar number density. Their weighted mean and standard deviation are $\Delta X_{\rm{C}}=-0\,.\!\!{^\prime}013\pm0\,.\!\!{^\prime}094$ and $\Delta Y_{\rm{C}}=-0\,.\!\!{^\prime}067\pm0\,.\!\!{^\prime}129$, respectively.

Given that our centers are derived from the stellar density in the regions outside the cluster core -- whereas the literature center is typically defined by the average coordinates of the brightest stars clustered in the innermost region -- the two definitions are not expected to coincide a priori. Our results show that the mean center of the fitted ellipses is offset by a distance of $\sim0\,.\!\!{^\prime}068$, or $4\,.\!\!{^{\prime\prime}}1\pm9\,.\!\!{^{\prime\prime}}6$. This represents an excellent agreement, with the uncertainty being on the order of the resolution of our density maps ($12^{\prime\prime}$--$24^{\prime\prime}$).

The results from the MCMC method show weighted mean offsets $\Delta X_{\rm{C}}=-0\,.\!\!{^\prime}069\pm0\,.\!\!{^\prime}056$ and $\Delta Y_{\rm{C}}=-0\,.\!\!{^\prime}071\pm0\,.\!\!{^\prime}023$ or the total offset is $5\,.\!\!{^{\prime\prime}}9\pm3\,.\!\!{^{\prime\prime}}6$ south-eastward from the adopted center value. Although the MCMC method provides a formally more precise determination of the  M\,15 central coordinates (as indicated by the smaller statistical uncertainties), its result shows a larger deviation from the literature value of \cite{Goldsbury+etal+2010}.
%This discrepancy suggests a potential systematic bias in our MCMC-based approach that is not captured by the formal errors.

In Figure~\ref{ell_a_all}, the results for the ellipticity are combined as a function of cluster-centric distance, $R$. Mean values are calculated for two regions: $R\leq4\,.\!\!{^\prime}5$ and $R>4\,.\!\!{^\prime}5$. This figure indicates a radial variation in the ellipticity of M\,15. The mean value is $\epsilon = 0.09\pm0.02$ for the inner region and $0.04\pm0.02$ for the outer region of the cluster. \cite{Bianchini+etal+2013} reveal that M\,15 exhibits a steep rotation profile that diminishes rapidly in the cluster's periphery. Specifically, systemic rotation becomes negligible at radial distances beyond 4--5 arcmin. This lack of rotational support likely accounts for the reduced ellipticity observed in the cluster's outskirts, where the morphology becomes increasingly spherical.

Crucially, this specific kinematic and morphological structure places
important constraints on the cluster's formation history. While standard
in-situ GCs are expected to lose significant angular
momentum through long-term dynamical relaxation, becoming largely
pressure-supported and spherical, M\,15 retains a rapidly rotating core.
Such a combination of high central rotation and strong central
concentration is physically consistent with expectations for the stripped
nuclei of dwarf galaxies, whose deep potential wells can preserve internal
rotation over long timescales. Therefore, our detection of a pronounced
ellipticity gradient from a rotationally-flattened core to a relaxed,
spherical halo provides morphological support for the hypothesis that M\,15
is the remnant nucleus of a disrupted satellite galaxy, rather than a
classical in-situ cluster. This evidence is particularly valuable given
the ambiguity in M\,15's current classification. We note that \cite{2024MNRAS.528.3198B} 
formally classify M\,15 as in-situ based on its orbital
energy and location within the central 10 kpc ($R_{\rm{GC}} \approx 10.4$ kpc, \citealt{Harris+1996}), but they explicitly flag it as a borderline
case due to its enhanced [Eu/Fe] abundance. This distinction is further
clarified by \cite{Monty+etal+2024}, whose analysis confirms that among the
other GC M\,15 shows the highest [Eu/$\alpha$] abundance, indicating a chemical
enrichment history distinct from that of the Milky Way. Consequently, our
structural analysis reinforces the scenario that M\,15 is an accreted nucleus that has 
settled into
the inner halo.

Figure~\ref{compare} compares our ellipticity results for $R>4\,.\!\!{^\prime}5$, derived from boxcar-smoothed stellar densities, with the findings of \cite{White+Shawl+1987}. Our results are in contradiction with theirs in the region $4\,.\!\!{^\prime}9<R<6^\prime$. Given that the errors on the ellipticities measured by \cite{White+Shawl+1987} are $\sim0.03$, this difference is most significant at $R=5\,.\!\!{^\prime}3$, reaching a level of $\sim3\sigma$. Our measured ellipticities are also inconsistent with the value from \cite{Chen+Chen+2010}, but a similar discrepancy was also found by \cite{Botev+etal+2025} for the GC M\,2. In contrast, our result is in excellent agreement with the finding of \cite{Reyes+Anderson+2024}, who reported $\epsilon=0.039 \pm 0.009$ at $R=7\,.\!\!{^\prime}5$.

We applied our ellipse fitting technique using recent Gaia EDR3 data provided by \cite{Vasiliev+Baumgardt+2021} and already utilized by \cite{Reyes+Anderson+2024}, in order to check whether the ellipticity varies with the distance from the center of M\,15. The sample contains 19770 member stars and we measured 67 ellipticities within the range of $3{^\prime}<R<7\,.\!\!{^\prime}5$ by fitting the smoothed number densities (see Figure \ref{compare}). It is evident that the variation of the ellipticity is in good agreement with our results based on PS1 DR2, confirming lower ellipticity in the outer regions ($R>4\,.\!\!{^\prime}5$) of M\,15 in comparison to the inner regions ($R<4\,.\!\!{^\prime}5$).

The MCMC analysis yields a weighted average ellipticity of $\epsilon = 0.022 \pm 0.005$ (see Figure~\ref{all_ell_mcmc}), which is representative of the entire cluster and agrees well with the ellipticities determined from ellipse fitting in the outer regions. However, considering the incompleteness of the samples in the central regions and the effects of stellar crowding, a single ellipticity value is likely insufficient to accurately represent the overall oblateness of the GC M\,15.

We find that the PA, as determined by the ellipse fitting of the stellar number density, can be assumed as a constant value: $\rm{PA} = 44\,.\!\!{\rm{^\circ}}4\pm16\,.\!\!{\rm{^\circ}}2$. The standard deviation is slightly affected by the weak radial trend of increasing PA with cluster-centric distance (see Figure~\ref{pa_a_all}).

The result from the MCMC method for the $\rm{PA} = 46\,.\!\!{\rm{^\circ}}6\pm7\,.\!\!{\rm{^\circ}}1$ (see Figure~\ref{all_ell_mcmc}) is in excellent agreement with the value from the ellipse fitting. \cite{Reyes+Anderson+2024} reported $\rm{PA} = 119\,.\!\!{\rm{^\circ}}2\pm19\,.\!\!{\rm{^\circ}}6$ at $R = 7\,.\!\!{^\prime}5$, which was measured with respect to West instead of North. Correcting for this offset, their value would be $29\,.\!\!{\rm{^\circ}}2\pm19\,.\!\!{\rm{^\circ}}6$, which is consistent with our results within 1$\sigma$.

The comparison of the structural parameters from this work and other studies is summarized in Table~\ref{Tab3} and Table~\ref{Tab4}.

%
%               one-column-spanning table
%________________________________________ Table 1: Use_of_the routines
\begin{table}
\begin{center}
\caption[]{Comparison of Position Angles Presented in This Work and Those Reported in Previous Studies.}\label{Tab3}

%%Please Capitalize the First Letter of Each Notional Word in table's caption

 \begin{tabular}{ccl}
  \hline\noalign{\smallskip}
  
   PA & PA err  & References\\
  $\rm{[deg]}$ & $\rm{[deg]}$ & \\
  \hline\noalign{\smallskip}

29.2 & 19.6 & \cite{Reyes+Anderson+2024}\\

36.1 & 2.9 & \cite{Martens+etal+2023}\\

76.6 & 12.1 & \cite{Sollima+etal+2019}\\

35 & 1 & \cite{White+Shawl+1987}\\

44.4 & 16.2 & present work (ellipse fitting)\\

46.6 & 7.1 & present work (MCMC method)\\

  \noalign{\smallskip}\hline
\end{tabular}
\end{center}
\end{table}

\begin{table}
\begin{center}
\caption[]{Comparison of Ellipticities $\epsilon$ from This Paper with Other Studies Measured at Different Distances $R$  from the GC Center.}\label{Tab4}

%%Please Capitalize the First Letter of Each Notional Word in table's caption

 \begin{tabular}{clcl}
  \hline\noalign{\smallskip}
  $\epsilon$ & $\epsilon$ $err$ & $R$ [$^\prime$]  & References                    \\
    
  \hline\noalign{\smallskip}
 
0.19 & 0.1 & - & \cite{Chen+Chen+2010}\\

0.039 & 0.009 & 7.5 &\cite{Reyes+Anderson+2024}\\
 
0.05 & 0.01 & 1.9--4.1 & \cite{White+Shawl+1987}\\
 
0.13 & 0.02 & 4.9--6.0 & \cite{White+Shawl+1987}\\

0.09 & 0.02 & 3.0--4.5 & present work (ellipse fitting)\\

0.04 & 0.02 & 4.5--8.0 & present work (ellipse fitting)\\

0.022 & 0.005 & - & present work (MCMC method)\\

  \noalign{\smallskip}\hline
\end{tabular}
\end{center}
\end{table}

\section{Conclusions}
\label{sect:conclusion}

This study of the M\,15 is based on 38 stellar density intervals, derived from photometric data from the PS1 DR2 catalog (\citealt{Flewelling+etal+2020}). Our study focuses on the estimation of the structural parameters (center coordinates, ellipticity, and PA) of this GC via ellipse fitting and a 2D Plummer profile fit to its overall discrete stellar density distribution, using a maximum likelihood approach.

By applying several specific filtering techniques, we confirm the findings of \cite{Botev+etal+2025} that the averaged 2D distributions of the smoothed number density yield more accurate results for the structural parameters of GCs than the discrete number density. We also find that the photometric quality flags serve as an important discriminator for the PS1 DR2 data samples. The results, based on the analysis of different photometric bands and angular scales, are in good agreement.

We find a radial variation in the ellipticity of M\,15, which decreases from a mean value of $\epsilon$ = $0.09\pm0.02$ at cluster-centric distances $R \leq 4\,.\!\!{^\prime}5$ to $0.04\pm0.02$ at $R > 4\,.\!\!{^\prime}5$ ($1\sigma$ confidence level). The latter value is in excellent agreement with the result of \cite{Reyes+Anderson+2024}, who found $\epsilon = 0.039\pm0.009$ at $R = 7\,.\!\!{^\prime}5$. Theoretically,  ellipticity is expected to increase in the outer region, because it is generally more susceptible to elongation due to tidal effects, while the inner region has a faster relaxation time, which acts towards decreasing anisotropy and increasing sphericity. However, the significant rotation of M\,15, characterized by the steep profile presented by \cite{Bianchini+etal+2013} (see their Figure 10), provides a mechanism for the observed core ellipticity. The relatively low ellipticity in the outer region suggests that there is no significant stellar tidal stripping in M\,15 at $R \sim 8^\prime$.

The results from the discrete MCMC fit show an ellipticity of $\epsilon = 0.022\pm0.005$, representative of the entire cluster, which also agrees well with the ellipticities determined from ellipse fitting in the outer regions. However, considering the incompleteness of the samples in the central regions and the effects of stellar crowding, a single ellipticity value is likely insufficient to accurately represent the overall oblateness of M\,15.

The mean offsets of the centers of all isodensity fitted ellipses are $\Delta\alpha=+0\,.\!\!{^\prime}013$ and $\Delta\delta =-0\,.\!\!{^\prime}067$, corresponding to a distance of $4\,.\!\!{^{\prime\prime}}1\pm9\,.\!\!{^{\prime\prime}}6$ from the previously precisely measured coordinates (\citealt{Goldsbury+etal+2010}) of the M\,15 center. This offset is on the order of the resolution of our maps ($12^{\prime\prime}$--$24^{\prime\prime}$).

We determine that the position angle, derived from the ellipse fitting of the stellar density, can be considered as nearly constant: $\rm{PA} = 44\,.\!\!{\rm{^\circ}}4\pm16\,.\!\!{\rm{^\circ}}2$. The result obtained from the MCMC method, $\rm{PA}=46\,.\!\!{\rm{^\circ}}6\pm7\,.\!\!{\rm{^\circ}}1$, is in excellent agreement with the ellipse fitting result.

\begin{acknowledgements}
The research that led to these results was partially carried out with
the help of infrastructure purchased under the National Roadmap for
Research Infrastructure, financially coordinated by the Ministry of
Education and Science of Republic of Bulgaria (D01-109/30.06.2025).

The Pan-STARRS1 Surveys (PS1) and the PS1 public science archive have been made possible through contributions by the Institute for Astronomy, the University of Hawaii, the Pan-STARRS Project Office, the Max-Planck Society and its participating institutes, the Max Planck Institute for Astronomy, Heidelberg and the Max Planck Institute for Extraterrestrial Physics, Garching, The Johns Hopkins University, Durham University, the University of Edinburgh, the Queen's University Belfast, the Harvard-Smithsonian Center for Astrophysics, the Las Cumbres Observatory Global Telescope Network Incorporated, the National Central University of Taiwan, the Space Telescope Science Institute, the National Aeronautics and Space Administration under Grant No. NNX08AR22G issued through the Planetary Science Division of the NASA Science Mission Directorate, the National Science Foundation Grant No. AST-1238877, the University of Maryland, Eotvos Lorand University (ELTE), the Los Alamos National Laboratory, and the Gordon and Betty Moore Foundation.
\end{acknowledgements}

\newpage

\appendix                  %%appendicial material is supported

\section{Supplementary Data}

This appendix presents the supplementary tables referenced in the main text. Table A1 contains the complete data obtained by ellipse fitting of the 2D stellar number isodensity distribution, while Table A2 lists the full derived structural parameters from MCMC.

\begin{table}[!htb]
\begin{center}
\caption[]{Ellipse Parameters Estimated from the Unfiltered $40^\prime$$\times$$40^\prime$ Sample under Varying Stellar Densities, Star Counting and Smoothing Techniques (see Section~\ref{iso} for details).}\label{Tab1}
\scalebox{0.77}{%
\begin{tabular}{c c c c c c c c c c c c c c c}
 \noalign{\smallskip}\hline
image size & boxcar & density & $X_{\rm{C}}$ & $err$ & $Y_{\rm{C}}$ & $err$ & $a$ & $err$ & $b$ & $err$ & PA & err & $\epsilon$ & $err$ \\ 
$\rm{[px}$$\times$$\rm{px]}$ & $\rm{[px}$$\times$$\rm{px]}$ & $[*/$$\rm{px}]$ & [ $^\prime$ ] & [ $^\prime$ ] & [ $^\prime$ ] & [ $^\prime$ ] & [ $^\prime$ ] & [ $^\prime$ ] & [ $^\prime$ ] & [ $^\prime$ ] & [deg] & [deg] & &  \\ 
 \noalign{\smallskip}\hline
100$\times$100 & - & 6 & -0.07 & 0.142 & -0.052 & 0.128 & 7.314 & 0.118 & 7.239 & 0.126 & 35.621 & 52.528 & 0.01 & 0.022 \\ 
  
100$\times$100 & - & 7-8 & 0.007 & 0.089 & 0.214 & 0.104 & 6.711 & 0.093 & 6.408 & 0.11 & 55.286 & 22.983 & 0.045 & 0.023 \\ 
  
100$\times$100 & - & 8-9 & -0.037 & 0.119 & 0.145 & 0.121 & 6.483 & 0.118 & 6.208 & 0.119 & 54.703 & 30.57 & 0.043 & 0.026 \\ 
  
100$\times$100 & - & 9-10 & -0.005 & 0.098 & -0.15 & 0.099 & 6.175 & 0.12 & 5.594 & 0.097 & 35.72 & 9.993 & 0.094 & 0.026 \\ 
  
100$\times$100 & - & 10-12 & 0.031 & 0.094 & -0.07 & 0.101 & 5.648 & 0.094 & 5.503 & 0.091 & 18.801 & 59.981 & 0.026 & 0.023 \\ 
  
100$\times$100 & - & 12-14 & 0.019 & 0.084 & 0.011 & 0.081 & 4.922 & 0.086 & 4.752 & 0.083 & 53.159 & 33.629 & 0.035 & 0.024 \\ 
  
100$\times$100 & - & 14-17 & -0.082 & 0.074 & -0.183 & 0.078 & 4.444 & 0.081 & 4.255 & 0.079 & 36.85 & 37.377 & 0.043 & 0.025 \\ 
  
100$\times$100 & 3$\times$3 & 5.0-6.0 & -0.045 & 0.042 & 0.018 & 0.045 & 7.334 & 0.056 & 7.047 & 0.049 & 25.22 & 13.564 & 0.039 & 0.011 \\ 
  
100$\times$100 & 3$\times$3 & 6.0-7.0 & -0.03 & 0.054 & -0.081 & 0.048 & 6.875 & 0.065 & 6.606 & 0.054 & 63.584 & 12.537 & 0.039 & 0.014 \\ 
  
100$\times$100 & 3$\times$3 & 7.0-8.0 & -0.026 & 0.047 & -0.054 & 0.049 & 6.431 & 0.053 & 6.268 & 0.053 & 46.806 & 24.169 & 0.025 & 0.013 \\ 
  
100$\times$100 & 3$\times$3 & 8.0-9.5 & -0.085 & 0.038 & -0.101 & 0.037 & 5.925 & 0.045 & 5.762 & 0.042 & 61.875 & 15.922 & 0.028 & 0.012 \\ 
  
100$\times$100 & 3$\times$3 & 9.5-12.0 & -0.006 & 0.032 & -0.118 & 0.03 & 5.387 & 0.037 & 5.205 & 0.037 & 53.736 & 10.375 & 0.034 & 0.011 \\ 
  
100$\times$100 & 3$\times$3 & 12.0-15.0 & -0.024 & 0.031 & -0.128 & 0.033 & 4.851 & 0.039 & 4.581 & 0.039 & 32.44 & 6.755 & 0.056 & 0.013 \\ 
  
100$\times$100 & 3$\times$3 & 15.0-18.0 & -0.057 & 0.04 & -0.157 & 0.038 & 4.329 & 0.041 & 3.876 & 0.049 & 21.395 & 4.925 & 0.105 & 0.016 \\ 
  
100$\times$100 & 5$\times$5 & 4.0-5.0 & -0.054 & 0.032 & 0.048 & 0.032 & 7.955 & 0.044 & 7.599 & 0.035 & 38.733 & 5.119 & 0.045 & 0.008 \\ 
  
100$\times$100 & 5$\times$5 & 5.0-6.0 & -0.041 & 0.033 & -0.037 & 0.033 & 7.307 & 0.041 & 7.094 & 0.036 & 51.853 & 9.871 & 0.029 & 0.008 \\ 
  
100$\times$100 & 5$\times$5 & 6.0-7.0 & -0.026 & 0.032 & -0.095 & 0.03 & 6.759 & 0.035 & 6.648 & 0.034 & 64.685 & 20.231 & 0.017 & 0.008 \\ 
  
100$\times$100 & 5$\times$5 & 7.0-9.0 & -0.024 & 0.029 & -0.145 & 0.027 & 6.194 & 0.035 & 6.045 & 0.031 & 63.336 & 11.843 & 0.024 & 0.009 \\ 
  
100$\times$100 & 5$\times$5 & 9.0-11.0 & -0.012 & 0.026 & -0.173 & 0.025 & 5.596 & 0.031 & 5.403 & 0.031 & 59.814 & 7.756 & 0.035 & 0.009 \\ 
  
100$\times$100 & 5$\times$5 & 11.0-14.0 & -0.01 & 0.028 & -0.136 & 0.028 & 5.002 & 0.034 & 4.763 & 0.036 & 40.688 & 6.758 & 0.048 & 0.011 \\ 
  
100$\times$100 & 5$\times$5 & 14.0-17.0 & -0.056 & 0.035 & -0.142 & 0.035 & 4.461 & 0.04 & 4.068 & 0.044 & 26.852 & 5.195 & 0.088 & 0.014 \\ 
  
200$\times$200 & - & 6 & -0.104 & 0.086 & -0.09 & 0.082 & 3.677 & 0.08 & 3.347 & 0.085 & 17.68 & 39.712 & 0.09 & 0.032 \\ 
  
200$\times$200 & - & 7 & 0.088 & 0.094 & 0.245 & 0.072 & 3.14 & 0.09 & 2.791 & 0.083 & 57.57 & 12.681 & 0.111 & 0.038 \\ 
  
200$\times$200 & - & 8-9 & 0.278 & 0.121 & -0.059 & 0.079 & 3.014 & 0.071 & 2.823 & 0.125 & 21.951 & 53.364 & 0.063 & 0.046 \\ 
  
200$\times$200 & 3$\times$3 & 2.0-2.1 & 0.027 & 0.071 & -0.06 & 0.064 & 6.359 & 0.072 & 6.214 & 0.064 & 60.173 & 30.108 & 0.023 & 0.015 \\ 
  
200$\times$200 & 3$\times$3 & 2.3-2.4 & 0.021 & 0.06 & -0.174 & 0.05 & 5.808 & 0.071 & 5.495 & 0.063 & 59.375 & 11.491 & 0.054 & 0.018 \\ 
  
200$\times$200 & 3$\times$3 & 2.6-2.8 & -0.035 & 0.049 & -0.203 & 0.056 & 5.489 & 0.07 & 5.149 & 0.052 & 41.934 & 9.217 & 0.062 & 0.017 \\ 
  
200$\times$200 & 3$\times$3 & 3.0-3.2 & 0.06 & 0.039 & -0.089 & 0.038 & 5.084 & 0.048 & 4.825 & 0.045 & 47.223 & 8.938 & 0.051 & 0.014 \\ 
  
200$\times$200 & 3$\times$3 & 3.4-3.6 & -0.091 & 0.041 & -0.084 & 0.049 & 4.862 & 0.057 & 4.48 & 0.053 & 23.446 & 10.155 & 0.079 & 0.018 \\ 
  
200$\times$200 & 3$\times$3 & 3.8-4.0 & -0.248 & 0.08 & -0.272 & 0.101 & 4.485 & 0.119 & 3.968 & 0.098 & 35.711 & 11.758 & 0.115 & 0.037 \\ 
  
200$\times$200 & 3$\times$3 & 4.2-4.4 & -0.011 & 0.051 & -0.271 & 0.047 & 4.122 & 0.05 & 3.763 & 0.061 & 38.808 & 8.14 & 0.087 & 0.02 \\ 
  
200$\times$200 & 5$\times$5 & 1.7-1.8 & -0.026 & 0.044 & -0.167 & 0.041 & 6.583 & 0.049 & 6.414 & 0.041 & 82.714 & 15.975 & 0.026 & 0.01 \\ 
  
200$\times$200 & 5$\times$5 & 2.0-2.1 & -0.036 & 0.042 & -0.113 & 0.04 & 6.198 & 0.05 & 5.907 & 0.044 & 66.495 & 8.322 & 0.047 & 0.012 \\ 
  
200$\times$200 & 5$\times$5 & 2.3-2.5 & 0.0 & 0.027 & -0.035 & 0.029 & 5.704 & 0.033 & 5.438 & 0.036 & 59.372 & 6.292 & 0.047 & 0.01 \\ 
  
200$\times$200 & 5$\times$5 & 2.7-2.9 & -0.02 & 0.024 & -0.12 & 0.025 & 5.275 & 0.035 & 5.108 & 0.031 & 43.554 & 9.18 & 0.032 & 0.011 \\ 
  
200$\times$200 & 5$\times$5 & 3.1-3.3 & -0.004 & 0.027 & -0.091 & 0.024 & 5.007 & 0.032 & 4.721 & 0.035 & 27.474 & 5.024 & 0.057 & 0.011 \\ 
  
200$\times$200 & 5$\times$5 & 3.5-3.7 & -0.016 & 0.028 & -0.159 & 0.024 & 4.61 & 0.027 & 4.425 & 0.035 & 40.082 & 8.276 & 0.04 & 0.012 \\ 
  
200$\times$200 & 5$\times$5 & 3.9-4.1 & -0.077 & 0.036 & -0.174 & 0.034 & 4.398 & 0.04 & 3.919 & 0.042 & 22.297 & 4.275 & 0.109 & 0.014 \\ 
  \noalign{\smallskip}\hline
\end{tabular}
}
\end{center}
\end{table}

\begin{table}
\begin{center}
\caption[]{MCMC Obtained Structural Parameters for Samples with Different Field Sizes, Filtering Conditions, and Background Subtraction Methods (see Section~\ref{MCMC} for details).}\label{Tab2}
\scalebox{0.77}{%
\begin{tabular}{c c c c c c c c c c c c}
  \hline\noalign{\smallskip}
image & filtration & background & $X_{\rm{C}}$ & $err$ & $Y_{\rm{C}}$  & $err$ &   $PA$&  $err$ & $\epsilon$ & $err$ &  Number    \\ 
 size & type & subtraction & [ $^\prime$ ] & [ $^\prime$ ] & [ $^\prime$ ] & [ $^\prime$ ] & [deg] & [deg] &  &  & of stars  \\ 
 \hline\noalign{\smallskip}
20$^\prime$$\times$20$^\prime$ 	&	 f0 	&	 b0 	&	-0.033	&	0.02	&	-0.117	&	0.019	&	33.53	&	6.131	&	0.035	&	0.008	&	18515	\\
20$^\prime$$\times$20$^\prime$ 	&	 f0 	&	 b1 	&	-0.125	&	0.024	&	-0.074	&	0.024	&	56.505	&	9.225	&	0.027	&	0.009	&	11278	\\
20$^\prime$$\times$20$^\prime$ 	&	 f0 	&	 b2 	&	-0.108	&	0.027	&	-0.092	&	0.029	&	54.385	&	9.626	&	0.030	&	0.010	&	10997	\\
20$^\prime$$\times$20$^\prime$ 	&	 f1 	&	 b0 	&	0.007	&	0.024	&	-0.099	&	0.022	&	38.514	&	10.886	&	0.023	&	0.008	&	14303	\\
20$^\prime$$\times$20$^\prime$ 	&	 f1 	&	 b1 	&	0.007	&	0.023	&	-0.085	&	0.026	&	45.848	&	9.855	&	0.024	&	0.008	&	13088	\\
20$^\prime$$\times$20$^\prime$ 	&	 f1 	&	 b2 	&	-0.004	&	0.023	&	-0.117	&	0.027	&	40.348	&	9.167	&	0.027	&	0.008	&	12845	\\
20$^\prime$$\times$20$^\prime$ 	&	 f2 	&	 b0 	&	-0.169	&	0.027	&	-0.081	&	0.027	&	53.01	&	9.74	&	0.028	&	0.009	&	13187	\\
20$^\prime$$\times$20$^\prime$ 	&	 f2 	&	 b1 	&	-0.156	&	0.026	&	-0.065	&	0.023	&	56.047	&	7.563	&	0.033	&	0.010	&	12028	\\
20$^\prime$$\times$20$^\prime$ 	&	 f2 	&	 b2 	&	-0.158	&	0.026	&	-0.087	&	0.025	&	52.208	&	7.964	&	0.031	&	0.008	&	11799	\\
20$^\prime$$\times$20$^\prime$ 	&	 f3 	&	 b0 	&	-0.044	&	0.025	&	-0.103	&	0.022	&	40.978	&	8.995	&	0.026	&	0.008	&	14919	\\
20$^\prime$$\times$20$^\prime$ 	&	 f3 	&	 b1 	&	-0.034	&	0.020	&	-0.084	&	0.022	&	46.822	&	7.277	&	0.031	&	0.008	&	13670	\\
20$^\prime$$\times$20$^\prime$ 	&	 f3 	&	 b2 	&	-0.051	&	0.021	&	-0.105	&	0.023	&	44.473	&	10.027	&	0.029	&	0.007	&	13414	\\
20$^\prime$$\times$20$^\prime$ 	&	 f4 	&	 b0 	&	-0.126	&	0.027	&	-0.082	&	0.023	&	49.573	&	12.949	&	0.019	&	0.009	&	12571	\\
20$^\prime$$\times$20$^\prime$ 	&	 f4 	&	 b1 	&	-0.107	&	0.026	&	-0.078	&	0.030	&	57.021	&	9.397	&	0.026	&	0.008	&	11444	\\
20$^\prime$$\times$20$^\prime$ 	&	 f4 	&	 b2 	&	-0.120	&	0.029	&	-0.101	&	0.028	&	54.213	&	9.568	&	0.027	&	0.008	&	11235	\\
30$^\prime$$\times$30$^\prime$ 	&	 f0 	&	 b0 	&	-0.032	&	0.019	&	-0.103	&	0.019	&	33.415	&	7.85	&	0.025	&	0.007	&	22332	\\
30$^\prime$$\times$30$^\prime$ 	&	 f0 	&	 b1 	&	-0.092	&	0.025	&	-0.053	&	0.025	&	49.974	&	11.688	&	0.023	&	0.008	&	12884	\\
30$^\prime$$\times$30$^\prime$ 	&	 f0 	&	 b2 	&	-0.087	&	0.031	&	-0.090	&	0.029	&	56.104	&	8.594	&	0.023	&	0.007	&	12201	\\
30$^\prime$$\times$30$^\prime$ 	&	 f1 	&	 b0 	&	0.002	&	0.02	&	-0.086	&	0.025	&	36.566	&	14.725	&	0.015	&	0.009	&	17793	\\
30$^\prime$$\times$30$^\prime$ 	&	 f1 	&	 b1 	&	-0.007	&	0.025	&	-0.074	&	0.024	&	48.598	&	9.110	&	0.021	&	0.007	&	14948	\\
30$^\prime$$\times$30$^\prime$ 	&	 f1 	&	 b2 	&	0.003	&	0.021	&	-0.097	&	0.020	&	44.301	&	9.053	&	0.021	&	0.007	&	14371	\\
30$^\prime$$\times$30$^\prime$ 	&	 f2 	&	 b0 	&	-0.156	&	0.03	&	-0.064	&	0.024	&	49.687	&	12.72	&	0.018	&	0.007	&	16612	\\
30$^\prime$$\times$30$^\prime$ 	&	 f2 	&	 b1 	&	-0.141	&	0.024	&	-0.041	&	0.025	&	58.568	&	9.053	&	0.024	&	0.006	&	13883	\\
30$^\prime$$\times$30$^\prime$ 	&	 f2 	&	 b2 	&	-0.125	&	0.023	&	-0.095	&	0.030	&	53.239	&	7.792	&	0.029	&	0.007	&	13320	\\
30$^\prime$$\times$30$^\prime$ 	&	 f3 	&	 b0 	&	-0.04	&	0.022	&	-0.079	&	0.023	&	38.171	&	9.11	&	0.022	&	0.007	&	18504	\\
30$^\prime$$\times$30$^\prime$ 	&	 f3 	&	 b1 	&	-0.040	&	0.021	&	-0.076	&	0.021	&	48.197	&	7.506	&	0.024	&	0.006	&	15591	\\
30$^\prime$$\times$30$^\prime$ 	&	 f3 	&	 b2 	&	-0.034	&	0.022	&	-0.096	&	0.023	&	47.510	&	7.792	&	0.027	&	0.007	&	14980	\\
30$^\prime$$\times$30$^\prime$ 	&	 f4 	&	 b0 	&	-0.12	&	0.026	&	-0.058	&	0.025	&	46.192	&	17.647	&	0.01	&	0.008	&	15901	\\
30$^\prime$$\times$30$^\prime$ 	&	 f4 	&	 b1 	&	-0.099	&	0.028	&	-0.059	&	0.023	&	55.188	&	8.537	&	0.023	&	0.007	&	13244	\\
30$^\prime$$\times$30$^\prime$ 	&	 f4 	&	 b2 	&	-0.088	&	0.027	&	-0.090	&	0.026	&	49.974	&	9.683	&	0.025	&	0.008	&	12706	\\
40$^\prime$$\times$40$^\prime$ 	&	 f0 	&	 b0 	&	-0.018	&	0.018	&	-0.106	&	0.016	&	37.025	&	7.907	&	0.023	&	0.007	&	25719	\\
40$^\prime$$\times$40$^\prime$ 	&	 f0 	&	 b1 	&	-0.013	&	0.019	&	-0.088	&	0.020	&	39.947	&	7.047	&	0.026	&	0.006	&	19802	\\
40$^\prime$$\times$40$^\prime$ 	&	 f0 	&	 b2 	&	-0.024	&	0.022	&	-0.107	&	0.022	&	41.207	&	6.245	&	0.028	&	0.006	&	18360	\\
40$^\prime$$\times$40$^\prime$ 	&	 f1 	&	 b0 	&	0.005	&	0.021	&	-0.077	&	0.024	&	33.358	&	13.121	&	0.017	&	0.009	&	20794	\\
40$^\prime$$\times$40$^\prime$ 	&	 f1 	&	 b1 	&	0.006	&	0.022	&	-0.050	&	0.021	&	45.848	&	10.829	&	0.020	&	0.007	&	15600	\\
40$^\prime$$\times$40$^\prime$ 	&	 f1 	&	 b2 	&	-0.004	&	0.023	&	-0.068	&	0.024	&	39.603	&	8.308	&	0.023	&	0.006	&	14576	\\
40$^\prime$$\times$40$^\prime$ 	&	 f2 	&	 b0 	&	-0.147	&	0.023	&	-0.067	&	0.026	&	49.573	&	9.912	&	0.018	&	0.008	&	19499	\\
40$^\prime$$\times$40$^\prime$ 	&	 f2 	&	 b1 	&	-0.142	&	0.026	&	-0.048	&	0.024	&	53.182	&	8.079	&	0.024	&	0.007	&	14501	\\
40$^\prime$$\times$40$^\prime$ 	&	 f2 	&	 b2 	&	-0.133	&	0.026	&	-0.069	&	0.026	&	51.062	&	8.709	&	0.026	&	0.007	&	13496	\\
40$^\prime$$\times$40$^\prime$ 	&	 f3 	&	 b0 	&	-0.036	&	0.024	&	-0.082	&	0.025	&	39.03	&	10.371	&	0.021	&	0.008	&	21596	\\
40$^\prime$$\times$40$^\prime$ 	&	 f3 	&	 b1 	&	-0.032	&	0.023	&	-0.049	&	0.022	&	50.203	&	8.652	&	0.024	&	0.007	&	16279	\\
40$^\prime$$\times$40$^\prime$ 	&	 f3 	&	 b2 	&	-0.044	&	0.024	&	-0.072	&	0.023	&	46.135	&	7.563	&	0.024	&	0.006	&	15196	\\
40$^\prime$$\times$40$^\prime$ 	&	 f4 	&	 b0 	&	-0.115	&	0.023	&	-0.058	&	0.03	&	47.854	&	14.152	&	0.013	&	0.008	&	18697	\\
40$^\prime$$\times$40$^\prime$ 	&	 f4 	&	 b1 	&	-0.093	&	0.025	&	-0.039	&	0.023	&	51.234	&	9.626	&	0.02	&	0.008	&	13820	\\
40$^\prime$$\times$40$^\prime$ 	&	 f4 	&	 b2 	&	-0.092	&	0.027	&	-0.061	&	0.024	&	49.229	&	8.594	&	0.025	&	0.007	&	12878	\\
50$^\prime$$\times$50$^\prime$ 	&	 f0 	&	 b0 	&	-0.024	&	0.017	&	-0.095	&	0.018	&	32.384	&	8.136	&	0.023	&	0.007	&	29757	\\
50$^\prime$$\times$50$^\prime$ 	&	 f0 	&	 b1 	&	-0.009	&	0.017	&	-0.079	&	0.021	&	39.89	&	7.735	&	0.023	&	0.006	&	20444	\\
50$^\prime$$\times$50$^\prime$ 	&	 f0 	&	 b2 	&	-0.019	&	0.02	&	-0.096	&	0.021	&	38.858	&	7.162	&	0.026	&	0.007	&	18188	\\
50$^\prime$$\times$50$^\prime$ 	&	 f1 	&	 b0 	&	0.006	&	0.023	&	-0.091	&	0.024	&	35.993	&	14.324	&	0.015	&	0.007	&	24362	\\
50$^\prime$$\times$50$^\prime$ 	&	 f1 	&	 b1 	&	0.013	&	0.023	&	-0.05	&	0.022	&	43.614	&	11.287	&	0.017	&	0.008	&	16150	\\
50$^\prime$$\times$50$^\prime$ 	&	 f1 	&	 b2 	&	-0.015	&	0.025	&	-0.061	&	0.025	&	41.322	&	10.371	&	0.019	&	0.007	&	14485	\\
50$^\prime$$\times$50$^\prime$ 	&	 f2 	&	 b0 	&	-0.148	&	0.024	&	-0.064	&	0.025	&	50.031	&	10.829	&	0.016	&	0.007	&	22895	\\
50$^\prime$$\times$50$^\prime$ 	&	 f2 	&	 b1 	&	-0.131	&	0.022	&	-0.03	&	0.025	&	55.188	&	6.818	&	0.027	&	0.007	&	15054	\\
50$^\prime$$\times$50$^\prime$ 	&	 f2 	&	 b2 	&	-0.132	&	0.027	&	-0.06	&	0.027	&	49.458	&	8.422	&	0.022	&	0.007	&	13394	\\
50$^\prime$$\times$50$^\prime$ 	&	 f3 	&	 b0 	&	-0.032	&	0.021	&	-0.078	&	0.022	&	37.941	&	12.433	&	0.017	&	0.008	&	25236	\\
50$^\prime$$\times$50$^\prime$ 	&	 f3 	&	 b1 	&	-0.028	&	0.022	&	-0.068	&	0.024	&	50.317	&	10.313	&	0.02	&	0.007	&	16851	\\
50$^\prime$$\times$50$^\prime$ 	&	 f3 	&	 b2 	&	-0.05	&	0.024	&	-0.067	&	0.023	&	46.327	&	10.159	&	0.021	&	0.007	&	15097	\\
50$^\prime$$\times$50$^\prime$ 	&	 f4 	&	 b0 	&	-0.109	&	0.025	&	-0.058	&	0.028	&	44.76	&	19.481	&	0.01	&	0.008	&	22021	\\
50$^\prime$$\times$50$^\prime$ 	&	 f4 	&	 b1 	&	-0.082	&	0.025	&	-0.031	&	0.025	&	56.849	&	8.48	&	0.025	&	0.007	&	14355	\\
50$^\prime$$\times$50$^\prime$ 	&	 f4 	&	 b2 	&	-0.095	&	0.027	&	-0.035	&	0.021	&	52.781	&	10.199	&	0.021	&	0.008	&	12784	\\
60$^\prime$$\times$60$^\prime$ 	&	 f0 	&	 b0 	&	-0.026	&	0.018	&	-0.103	&	0.021	&	34.045	&	7.678	&	0.022	&	0.007	&	34502	\\
60$^\prime$$\times$60$^\prime$ 	&	 f0 	&	 b1 	&	-0.011	&	0.02	&	-0.088	&	0.02	&	44.301	&	7.678	&	0.023	&	0.006	&	20895	\\
60$^\prime$$\times$60$^\prime$ 	&	 f0 	&	 b2 	&	-0.011	&	0.022	&	-0.075	&	0.02	&	39.775	&	7.62	&	0.025	&	0.006	&	17577	\\
60$^\prime$$\times$60$^\prime$ 	&	 f1 	&	 b0 	&	0.01	&	0.022	&	-0.079	&	0.022	&	32.613	&	13.866	&	0.014	&	0.008	&	28570	\\
60$^\prime$$\times$60$^\prime$ 	&	 f1 	&	 b1 	&	0.009	&	0.025	&	-0.056	&	0.022	&	47.567	&	16.616	&	0.011	&	0.008	&	16569	\\
60$^\prime$$\times$60$^\prime$ 	&	 f1 	&	 b2 	&	0.005	&	0.025	&	-0.046	&	0.022	&	41.207	&	10.485	&	0.02	&	0.007	&	14064	\\
60$^\prime$$\times$60$^\prime$ 	&	 f2 	&	 b0 	&	-0.148	&	0.023	&	-0.066	&	0.022	&	50.26	&	12.605	&	0.016	&	0.008	&	26906	\\
60$^\prime$$\times$60$^\prime$ 	&	 f2 	&	 b1 	&	-0.139	&	0.028	&	-0.019	&	0.025	&	57.709	&	9.11	&	0.021	&	0.008	&	15439	\\
60$^\prime$$\times$60$^\prime$ 	&	 f2 	&	 b2 	&	-0.118	&	0.027	&	-0.031	&	0.022	&	51.177	&	9.912	&	0.022	&	0.008	&	12957	\\
60$^\prime$$\times$60$^\prime$ 	&	 f3 	&	 b0 	&	-0.032	&	0.023	&	-0.083	&	0.021	&	39.89	&	9.912	&	0.019	&	0.008	&	29532	\\
60$^\prime$$\times$60$^\prime$ 	&	 f3 	&	 b1 	&	-0.037	&	0.018	&	-0.051	&	0.022	&	49.057	&	12.662	&	0.014	&	0.008	&	17281	\\
60$^\prime$$\times$60$^\prime$ 	&	 f3 	&	 b2 	&	-0.027	&	0.026	&	-0.047	&	0.024	&	44.244	&	9.053	&	0.022	&	0.008	&	14669	\\
60$^\prime$$\times$60$^\prime$ 	&	 f4 	&	 b0 	&	-0.108	&	0.03	&	-0.064	&	0.025	&	49.573	&	14.439	&	0.013	&	0.008	&	25944	\\
60$^\prime$$\times$60$^\prime$ 	&	 f4 	&	 b1 	&	-0.102	&	0.027	&	-0.021	&	0.021	&	57.479	&	11.803	&	0.015	&	0.008	&	14709	\\
60$^\prime$$\times$60$^\prime$ 	&	 f4 	&	 b2 	&	-0.086	&	0.027	&	-0.028	&	0.029	&	54.328	&	10.027	&	0.021	&	0.008	&	12366	\\
\noalign{\smallskip}\hline

\end{tabular}
}
\end{center}
\end{table}

\newpage

\section{Testing the Effects of Boxcar Smoothing on Ellipticity Measurements}
\label{AppB}
\begin{figure}[h]
\centering
\includegraphics[width=1\textwidth, angle=0]{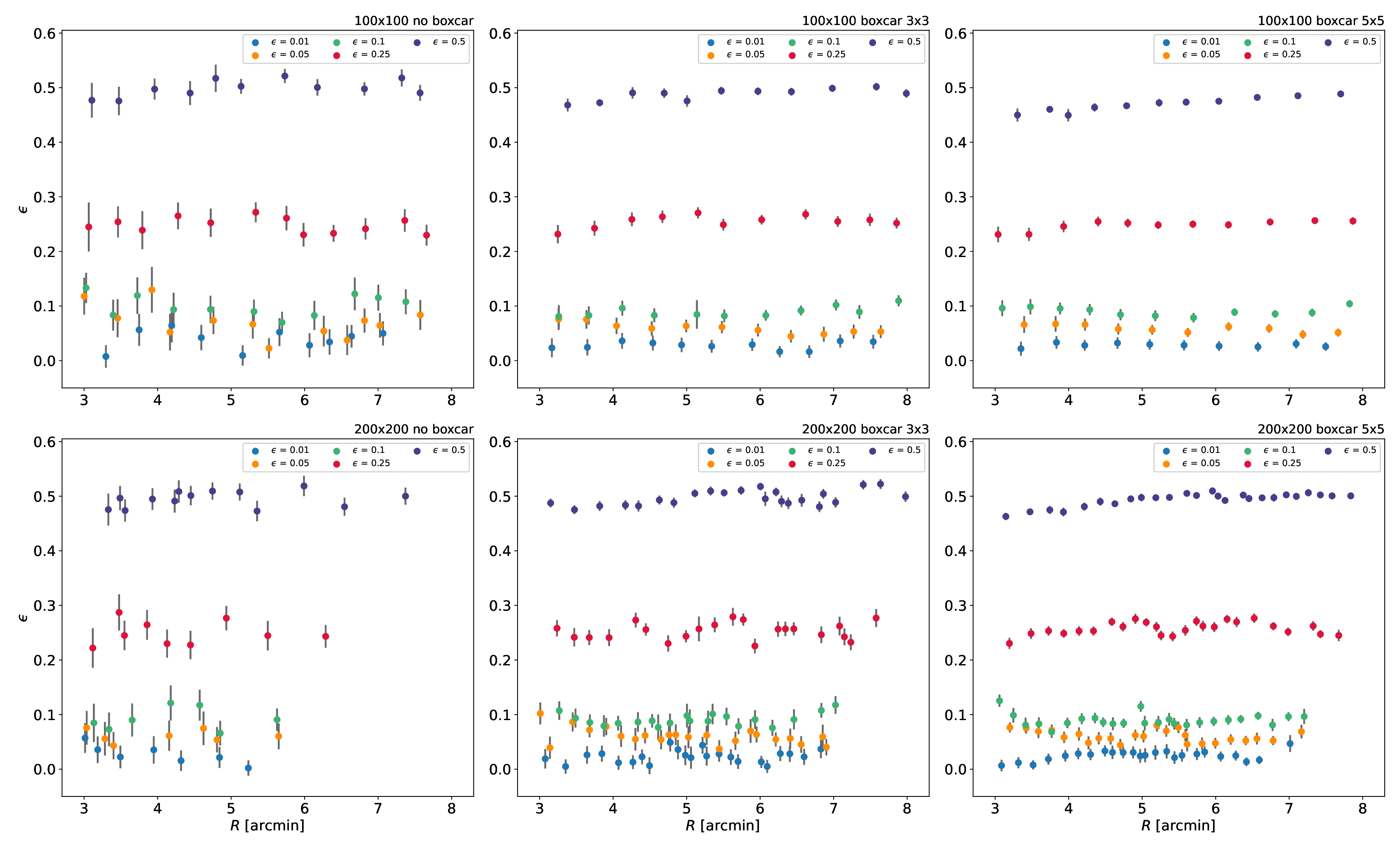}
\caption{Ellipticities derived from mock stellar density maps as a function of radius $R$. The mock maps were generated with ellipticities 0.01 (blue), 0.05 (orange), 0.1 (green), 0.25 (red), 0.5 (purple) and a constant PA = 45$^\circ$ along the radial distance $R$. From left to right the first row presents the 100×100 array under the three conditions – unsmoothed (left), 3×3 boxcar (middle) and 5×5 boxcar (right). The second row shows the same arrangement for the 200×200 array.}
\label{all_ell_test}
\end{figure}

To ensure that the pixel-binning and subsequent boxcar smoothing procedure utilized in our methodology does not introduce any systematic bias into the derived structural parameters, we performed a series of validation tests using mock stellar density maps. Following the approach used for the observational data, we generated synthetic 2D Plummer number density distributions with a predefined range of constant ellipticities, covering different input values in the interval from $0.01$ to $0.5$. These mock images were constructed to simulate the spatial resolution and typical noise characteristics of our actual density maps. We then analyzed these mock images with our ellipse-fitting algorithm in two ways: first on the raw, unsmoothed pixel-binned maps, and second on the maps after applying the exact same boxcar smoothing kernel described in Section~\ref{iso}. The results of this validation are presented in Figure ~\ref{all_ell_test}. The two leftmost panels display the measured ellipticities obtained from the unsmoothed mock maps as a function of radius $R$. As expected, the inherent pixel noise introduces a noticeable scatter around the true input values. Conversely, the middle and right panels show the measurements obtained after applying our boxcar smoothing kernel. It is clearly visible that the smoothing process significantly reduces the scatter, aligning the measured ellipticities with the true input values. Therefore, we conclude that the applied boxcar smoothing process effectively reduces local pixel noise and improves the precision of the measurement, without artificially dampening, enhancing, or otherwise biasing the ellipticity of the cluster's isodensity contours.

\label{lastpage}

\end{document}